\documentclass[11pt,a4paper]{article}
\usepackage{a4wide}

\usepackage[utf8]{inputenc}

\usepackage{cite}

\usepackage{nameref,hyperref}

\usepackage[right]{lineno}

\usepackage{amsmath}
\usepackage{amsfonts}

\usepackage{microtype}
\DisableLigatures[f]{encoding = *, family = * }

\usepackage{graphicx}

\begin{document}
\vspace*{0.35in}

\begin{flushleft}
{\Large
\textbf{Performance of the CMS Zero Degree Calorimeters in pPb collisions at the LHC}
}
\newline
\\
O. Surányi\textsuperscript{1,*},
A.~Al-Bataineh\textsuperscript{3},
J.~Bowen\textsuperscript{3},
S.~Cooper\textsuperscript{7},
M.~Csan\'ad\textsuperscript{1},
V.~Hagopian\textsuperscript{6},
D.~Ingram\textsuperscript{5},
C.~Ferraioli\textsuperscript{4},
T.~Grassi\textsuperscript{4},
R.~Kellogg\textsuperscript{4},
E.~Laird\textsuperscript{9},
G.~Martinez\textsuperscript{6},
W.~McBrayer\textsuperscript{3},
A.~Mestvirishvili\textsuperscript{5},
A.~Mignerey\textsuperscript{4},
M.~Murray\textsuperscript{3},
M.~Nagy\textsuperscript{1},
Y.~Onel\textsuperscript{5},
F.~Sikl\'er\textsuperscript{2},
M.~Toms\textsuperscript{8},
G.~Veres\textsuperscript{1},
Q.~Wang\textsuperscript{3}
\\
\bigskip
\textsuperscript{1} MTA-ELTE Lend\"ulet CMS Particle and Nuclear Physics Group, E\"otv\"os Lor\'and University, Budapest, Hungary  \\
\textsuperscript{2} Wigner RCP, Budapest, Hungary  \\
\textsuperscript{3} University of Kansas, Lawrence, USA  \\
\textsuperscript{4} University of Maryland, College Park, USA  \\
\textsuperscript{5} University of Iowa, Iowa City, USA  \\
\textsuperscript{6} Florida State University, Tallahassee, USA  \\
\textsuperscript{7} University of Alabama, Tuscaloosa, USA  \\
\textsuperscript{8} NRC Kurchatov Institute (ITEP), Moscow, Russia  \\
\textsuperscript{9} Brown University, Providence, USA
\\
\bigskip
* oliver.suranyi@cern.ch

\end{flushleft}

\section*{Abstract}
The two Zero Degree Calorimeters (ZDCs) of the CMS experiment are located at $\pm 140~$m from the collision point and detect neutral particles in the $|\eta| > 8.3$ pseudorapidity region. This paper presents a study on the performance of the ZDC in the 2016 pPb run. The response of the detectors to ultrarelativistic neutrons is studied using in-depth Monte Carlo simulations. A method of signal extraction based on template fits is presented, along with a dedicated calibration procedure. A deconvolution technique for the correction of overlapping collision events is discussed.


\section{Introduction}
Many measurements involving proton-ion and heavy-ion collisions require the knowledge of the centrality of the collision \cite{Chatrchyan:2012mb,Adam:2014qja}. One way to determine this is by measuring the number of nucleons that do not participate in the collision. The SPS, RHIC, and LHC heavy-ion experiments have measured these spectator nucleons with Zero Degree Calorimeters (ZDCs). The CMS ZDCs are two identical forward calorimeters located between the two LHC beam pipes at a distance of approximately 140~m from the CMS interaction point along the beamline, on each side. There are numerous other applications of ZDC detectors, such as minimum bias triggering, study of ultraperipheral collisions, and charge exchange processes. This paper presents results demonstrating the performance of the ZDCs in the 2016 pPb data-taking period. The results presented here are based on a sample of 10 million minimum bias events collected at a center-of-mass energy of $\sqrt{s_{\mathrm{NN}}} = 8.16$~TeV. The ZDC detects the neutral fragments of the Pb ions, and the neutrons emitted from the ions are nearly monoenergetic, thus they provide a unique opportunity to study the performance of the detector. The paper first introduces the structure of the ZDC detectors, then a Monte Carlo simulation study of the behaviour of the detector is presented. Afterwards the signal extraction and the calibration process is discussed. Finally, a method based on Fourier transformation is presented to correct the measured spectrum for pileup collisions.

\section{The CMS Zero Degree Calorimeter}
\label{sec:zdc}
The ZDCs of the CMS experiment complement the main CMS detector especially for heavy ion studies. They reside in special detector slots in the neutral particle absorber (TAN), which protects the first superconducting quadrupole magnet from radiation. A full description of the ZDC can be found in \cite{Grachov:2006ke,Grachov:2007wh,Grachov:2008qg,Grachov:2010th}. Located inside the TAN at pseudorapidity $\eta$ greater than 8.3 corresponding roughly to $\theta < 0.5$~mrad, the ZDCs detect photons and those neutral particles that are not swept away by bending and focusing magnets between the interaction point and the ZDCs.

\begin{figure}[t]
\centering
\includegraphics[width=0.55\textwidth]{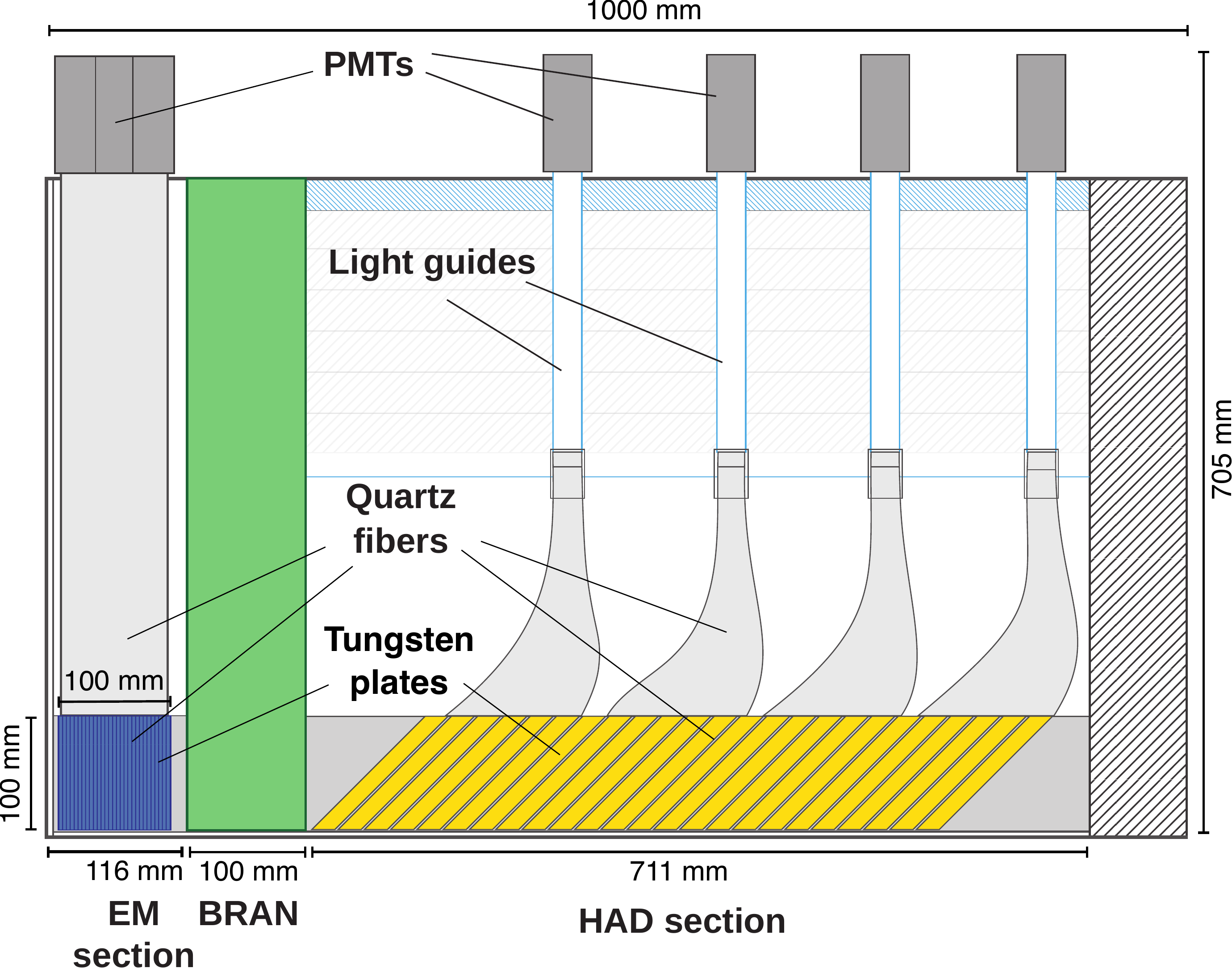}
\includegraphics[width=0.44\textwidth]{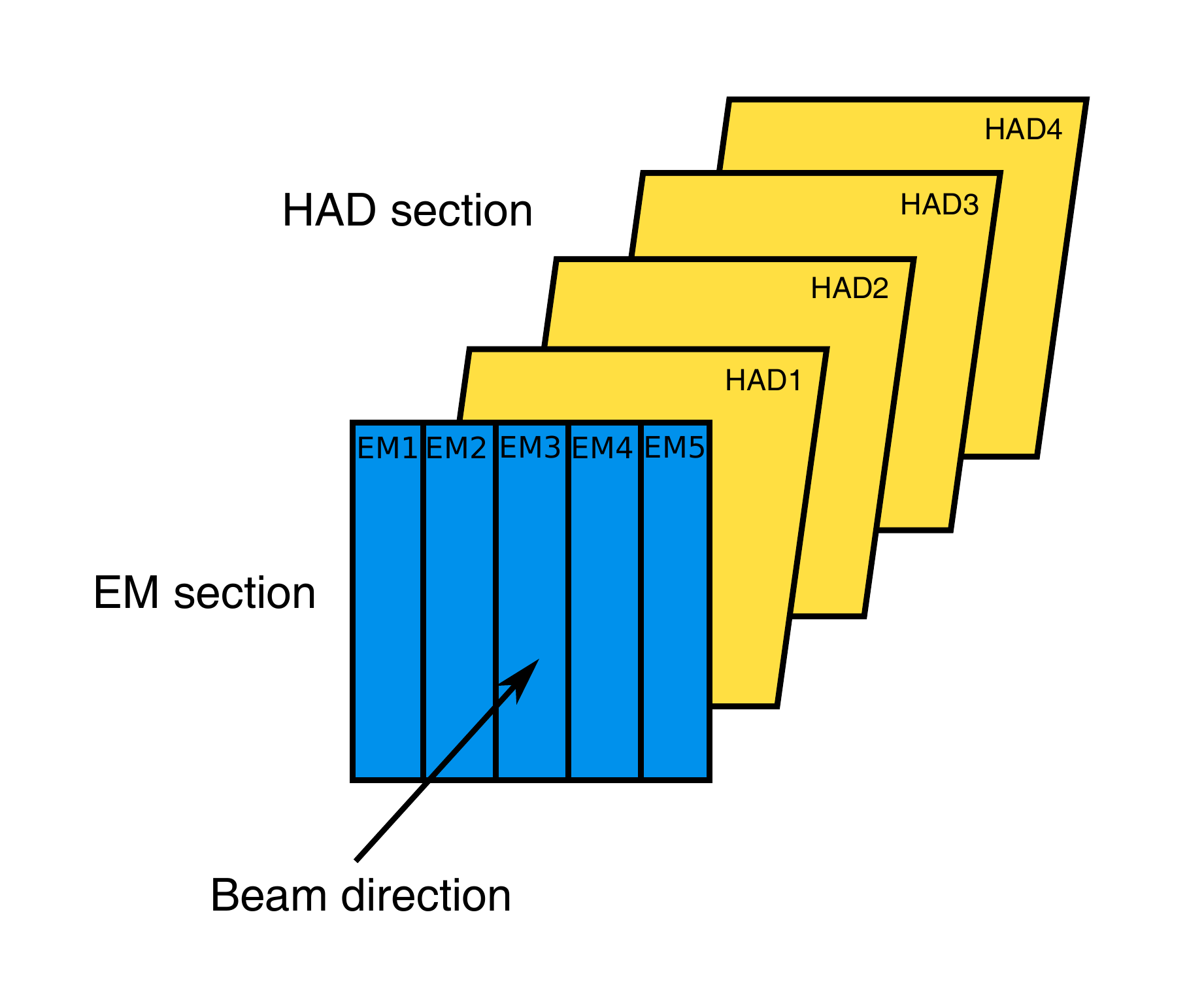}
\caption{The schematic side-view (left) and segmentation (right) of the CMS ZDC.}
\label{fig:zdc}
\end{figure}

\begin{table}[b]
\centering
\caption{Basic properties of the electromagnetic and hadronic sections of ZDC.}
\begin{tabular}{lcc}
\hline
& \multicolumn{1}{c}{Electromagnetic section} & \multicolumn{1}{c}{Hadronic section} \\
\hline
Sampling ratio &  2~mm W/0.7~mm QF & 15.5~mm W/0.7~mm QF \\
Number of cells & 33 & 24 \\
Radiaton/interaction length & $19X_0$ & $5.6\lambda_0$ \\
Number of channels & 5 horizontal divisions & 4 longitudinal segments  \\
Module size ($\text{W} \times \text{L} \times \text{H}$), mm & $92 \times 116 \times 705$ & $92 \times 711 \times 705$ \\
Weight of module, kg & $\approx 65$ & $\approx 400$ \\
\hline
\end{tabular}
\label{tab:zdc}
\end{table}

The schematic view of a ZDC detector is shown in the left panel of Fig.\ \ref{fig:zdc}. Each ZDC is a sampling calorimeter with tungsten absorber plates and quartz fibers (QF) as the active medium. The quartz fibers are routed to overhead photomultipliers. Each ZDC has two sections: an electromagnetic (EM) section optimized for photon detection and energy measurement, and a hadronic (HAD) section for measuring the energies of long-lived neutral hadrons. These neutral hadrons are dominantly neutrons, but K$^0_{\mathrm{L}}$ and $\Lambda^0$ particles can also reach the ZDCs. The basic properties of the EM and HAD sections are summarized in Table \ref{tab:zdc}. The EM section is segmented into 5 vertical strips that allows the determination of the horizontal position of the incoming particles. Tungsten plates and fibers in the EM section run vertically. The hadronic section is divided into 4 segments, as seen in Fig.\ \ref{fig:zdc}. In the hadronic section, the tungsten plates are tilted by $45^{\circ}$ to optimize the collection of Cherenkov light. The quartz fibers are clad in doped quartz, yielding a numerical aperature of 0.22. Individual fiber ribbons are grouped together to form a readout bundle that is compressed and glued into a circular shape. A light guide carries the light through radiation shielding to Hamamatsu R7525 photomultiplier tubes. Between the two ZDC sections lies an ionization chamber called BRAN (Beam RAte for Neutrals), which gives a measurement of the instantaneous luminosity which is independent of the operation of CMS \cite{Matis:2016raz}.

For each collision event, the signal is collected over 10 timeslices (TS) of 25~ns each. The peak of the ZDC signal always occurs in TS3, whereas due to the 100~ns bunch spacing, further signals may be present four timeslices before and after the main signal. The main signal is extracted from TS3. The high voltage powering the PMTs is set such that the analog-digital converters may saturate for larger signals. In this case, information from the tail of the signal is used to determine the total signal value. This preserves the excellent few neutron resolution if the number of neutrons is low (for example in an ultra-peripheral collision), while at the same time allowing the entire range of the ZDC to be exploited for centrality measurement using the tail of the signal.

\section{Monte Carlo modelling of the CMS ZDC}
\label{sec:mc}
The full ZDC geometry, including the BRAN detector, is modeled within the \textsc{Geant}4 framework (version 10.00.p03) \cite{Agostinelli:2002hh}. First the behavior of the detector is studied using monoenergetic neutrons parallel to the beamline. The Cherenkov photons produced by charged particles in the showers are generated in each simulation step \cite{Frank:1937fk}. The optical photons generated this way are required to fulfill the light guiding condition: their incident angle on the inner surface of the fibers should be larger than the corresponding critical angle for total internal reflection. Additionally the photons may be rejected based on the quantum efficiency of the PMTs. 


\begin{figure}[t]
\centering
\includegraphics[width=0.49\textwidth]{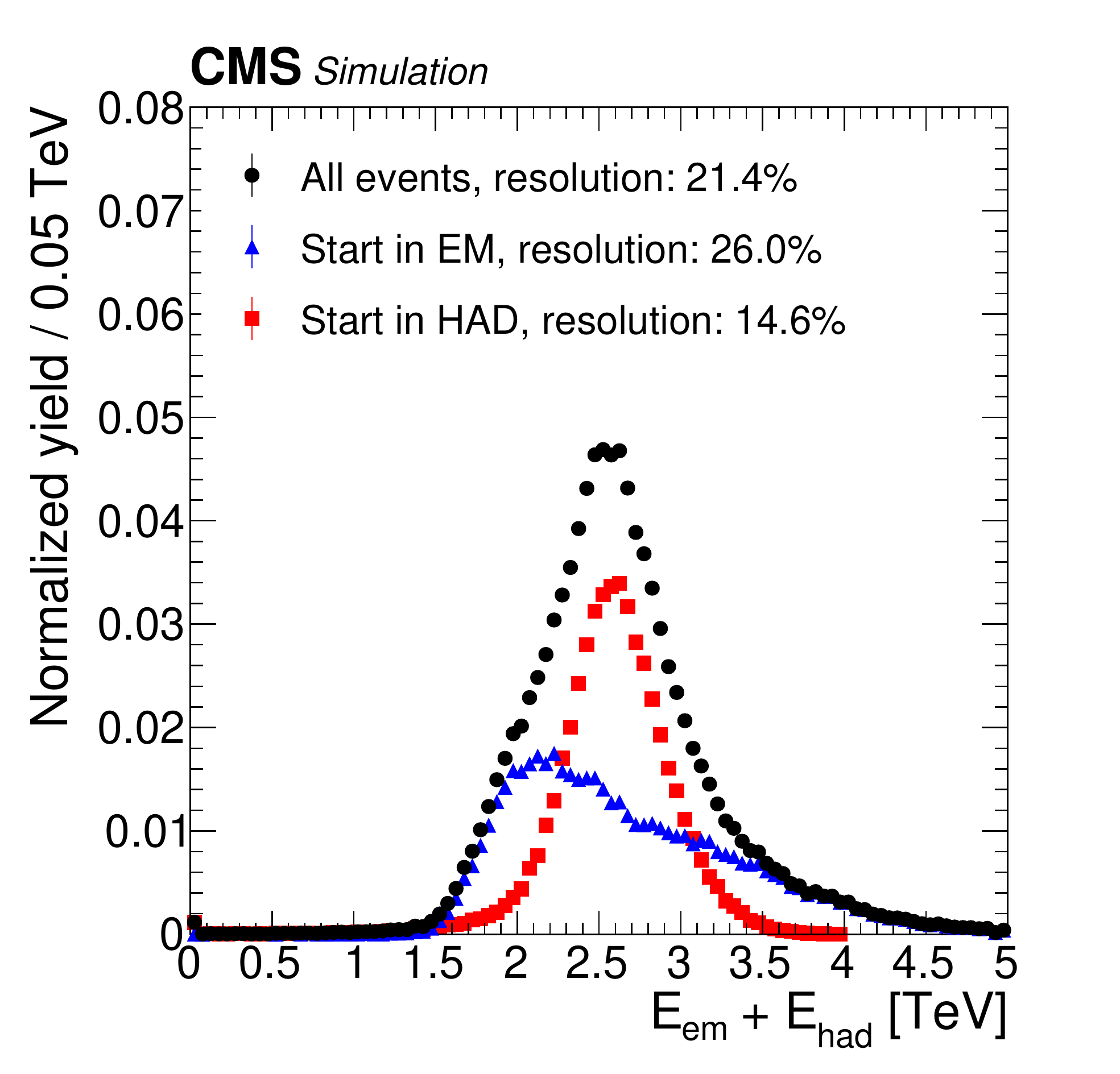}
\includegraphics[width=0.49\textwidth]{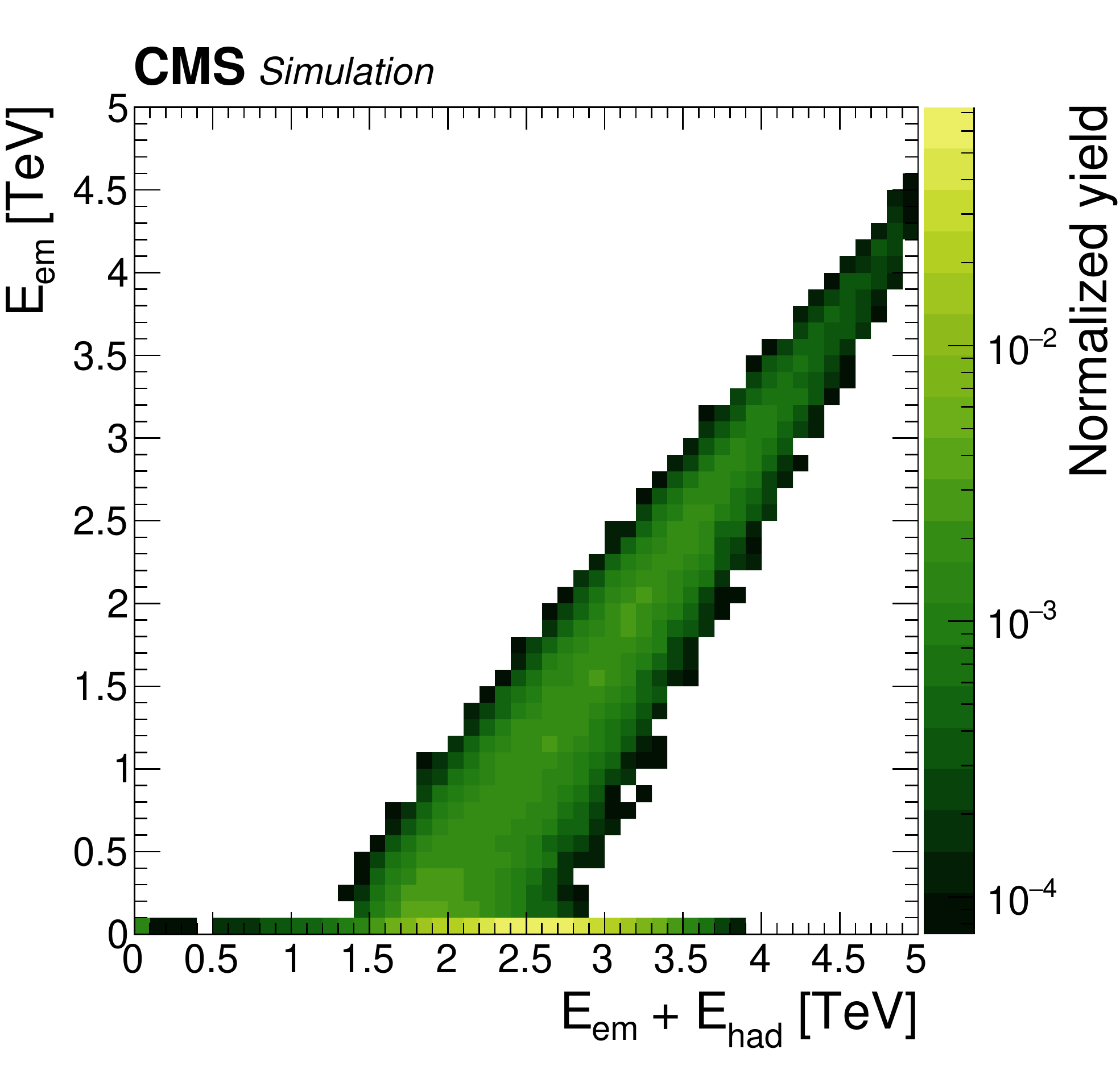}
\caption{The ZDC response for 2.56~TeV energy neutrons, separately plotted for neutrons which start to shower in the EM section and the HAD section (left), and the dependence of the response on energy deposited in the EM section (right).}
\label{fig:response1}
\end{figure}

\begin{figure}[t]
\centering
\includegraphics[width=0.49\textwidth]{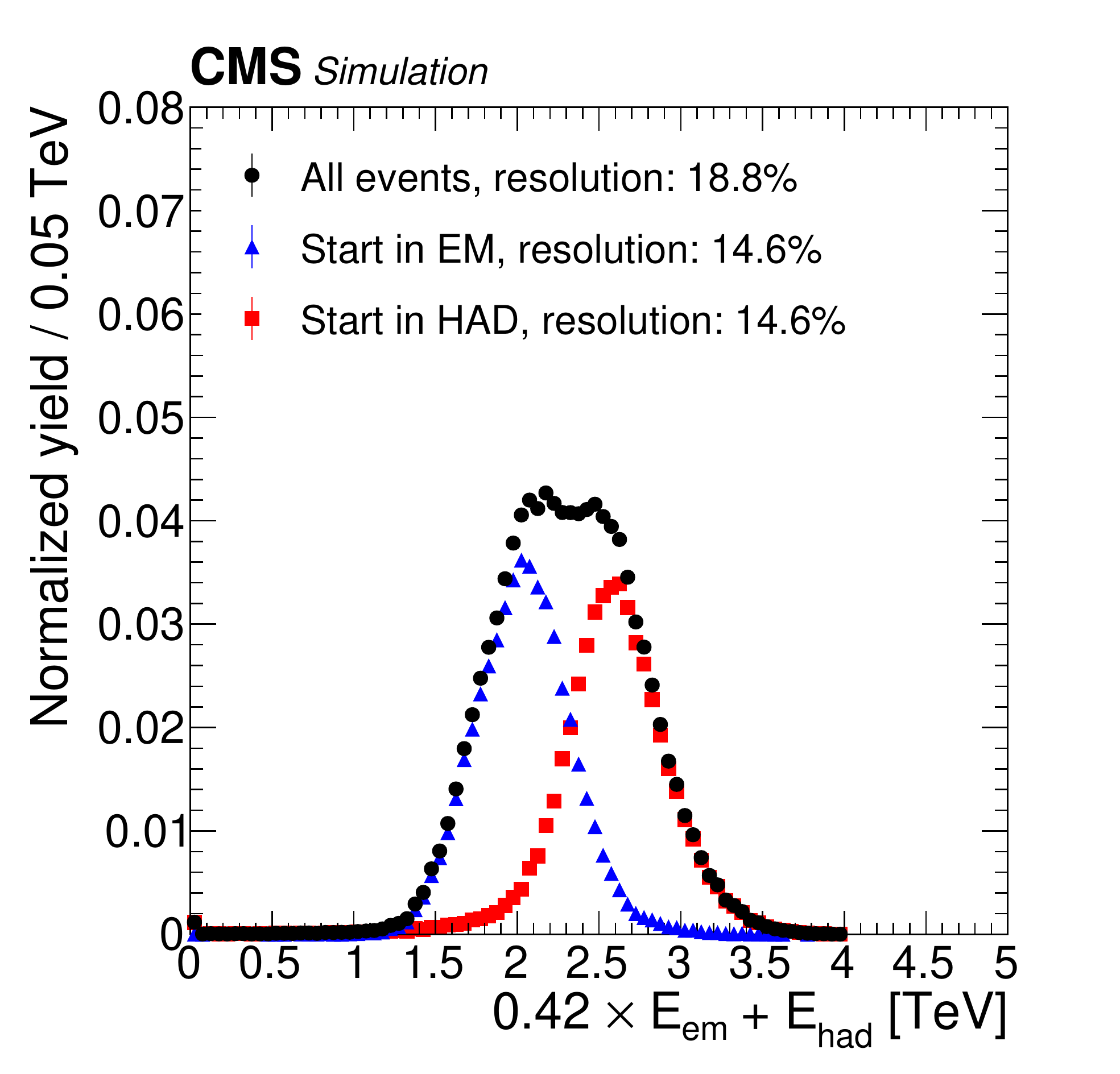}
\includegraphics[width=0.49\textwidth]{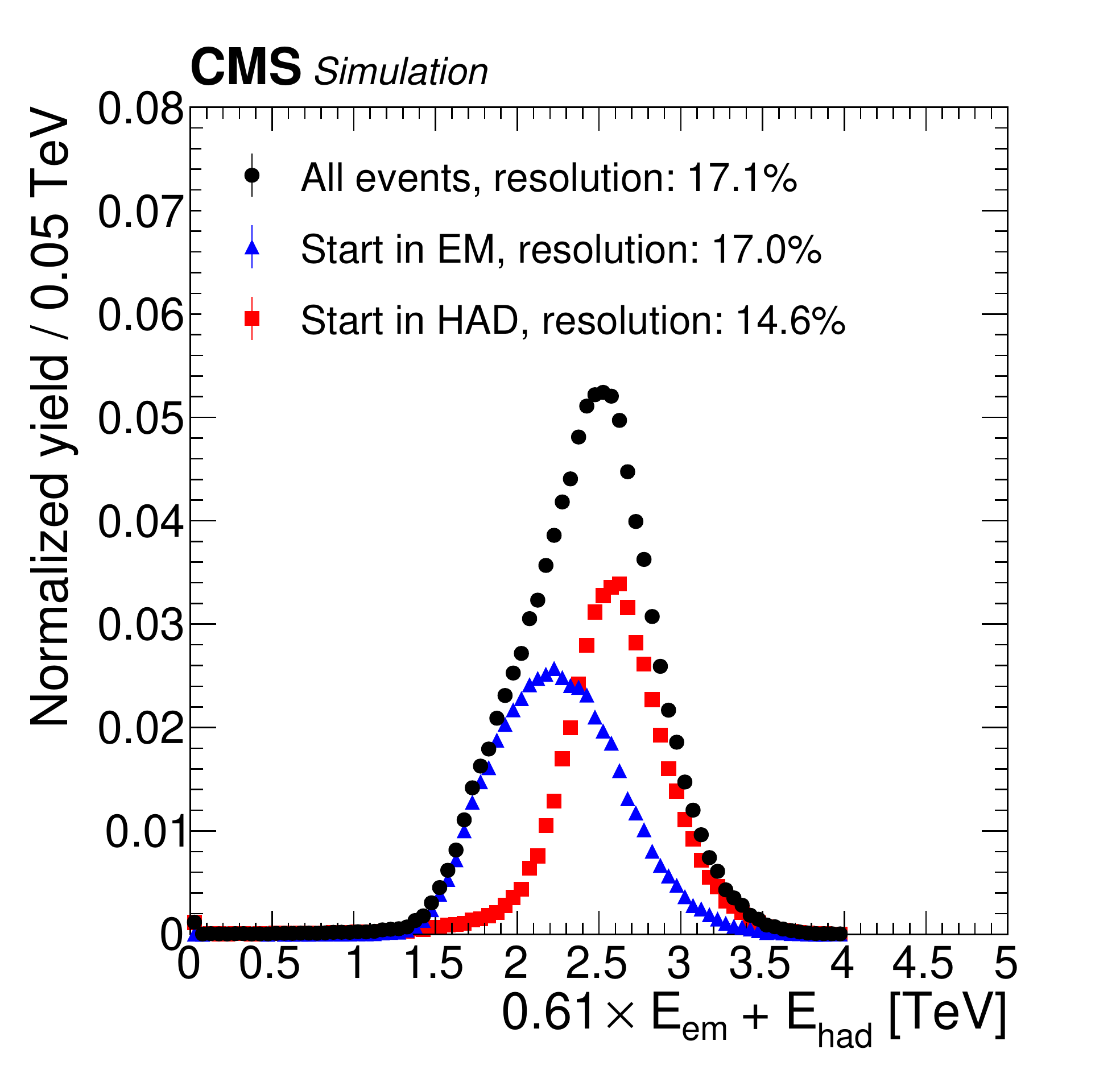}
\caption{The ZDC response for 2.56~TeV energy neutrons, with all EM energy deposits weighted by $w_{\text{EM}} = 0.42$ (left) and $0.61$ (right), which were calculated by minimizing the resolution using events that shower in the EM section and all events respectively.}
\label{fig:response2}
\end{figure}

The simulated ZDC response for 2.56~TeV monoenergetic neutrons is shown in the left panel of Fig.\ \ref{fig:response1} separately for neutrons which started showering either in the EM section or in the HAD section. The neutrons showering in the EM section have a much worse energy resolution than the ones showering only in the HAD section. The reason behind this is demonstrated in the right panel of Fig.\ \ref{fig:response1}: the measured energy of the neutrons depends on the fraction of their energy deposited in the EM section, since the total width of sensitive quartz fiber layers with respect to the tungsten absorber is higher in the EM section, therefore it samples a higher fraction of a shower. This difference can be corrected by multiplying all energy deposits in the EM channels with a $w_{\text{EM}}$ weighting factor. This factor is calculated by minimizing the relative energy resolution, defined as the ratio of the standard deviation and the mean of the measured energy values. First this calculation is performed using only the events with the shower starting in the EM section. The best resolution is achieved by using $w_{\text{EM}} = 0.42$, and the corresponding energy distributions are shown in the left panel of Fig.\ \ref{fig:response2}. In the corrected ZDC response, it is found that there is a shift between the two peaks as showers starting in the EM section are partly absorbed by the material of the BRAN detector. By the comparison of the position of the peaks it is concluded that approximately an average of 20\% of the energy is lost in those events that start to shower in the EM section. In case of real pPb collisions, in most cases more than one neutron is produced. Some of them may start showering already in the EM section, whereas others have the first interaction only in the HAD section. Therefore it is not possible to treat these two cases separately as they will be inevitably mixed, when in a real collision several neutrons hit the ZDC simultaneously. Alternatively, it is also possible to minimize the total resolution, resulting in $w_{\text{EM}} = 0.61$. The corresponding energy distribution is shown in the right panel of Fig.\ \ref{fig:response2}. This is not the optimal factor for the neutrons which shower in the EM, but this is the best overall resolution which can be achieved with the detector.

The signal detected in ZDC is dominated by neutrons emitted from the colliding Pb nuclei. The three main sources of these neutrons are nuclear evaporation processes \cite{Ferrari:1995cq,Sikler:2003ef}, intranuclear cascades \cite{Ferrari:1995cq,Sikler:2003ef} and neutrons emitted due to electromagnetic nuclear excitations such as the giant dipole resonances (GDR) \cite{RevModPhys.47.713,Pshenichnov:2011zz,Chiu:2001ij}. These neutrons are simulated in order to study ZDC acceptance and response. It is assumed that the neutrons are emitted isotropically in the rest frame of the nucleus according to the Maxwell-Boltzmann momentum distribution:
\begin{linenomath}
\begin{align}
\frac{\mathrm{d}N}{\mathrm{d}p} \propto p^2\exp\left(-\dfrac{p^2}{2m_nT}\right),
\end{align}
\end{linenomath}
where $p$ is the total momentum, $m_n$ is the neutron mass and $T$ is the Maxwell-Boltzmann temperature. The values of the $T$ parameter are 1, 5, and 50~MeV for neutrons originating from electromagnetic excitation \cite{GAYTHER1967733}, evaporation \cite{Sikler:2003ef} and intranuclear cascade \cite{Sikler:2003ef} processes respectively. All neutrons are boosted in the z-direction by $\gamma = 2752$, which is the Lorentz-factor of the Pb ion in pPb collisions at $\sqrt{s_{NN}} = 8.16$~TeV.

The effect of crossing angle, beam divergence and the smearing of the beamspot are taken into account by applying the following procedure on all generated neutrons. First the location of the interaction point $(v_x,v_y,v_z)$ is sampled from a Gaussian beamspot with position $(x_0,y_0,z_0)$ and size $(\sigma_x,\sigma_y,\sigma_z)$. The effect of beam divergence is taken into account by introducing the $M(z)$ magnification factor:
\begin{linenomath}
\begin{align}
M(z) = \frac{\sigma(z)}{\sigma(0)} = \sqrt{\frac{\varepsilon \beta(z)}{\varepsilon\beta^*}} = \sqrt{1 + \frac{z^2}{\beta^{*2}}},
\end{align}
\end{linenomath}
where $\sigma(z)$ is the transverse size of the beam at distance $z$ from the interaction point if no focusing is used, $\varepsilon$ is the beam emittance and $\beta(z)$ is the beta-function, with $\beta(0) = \beta^*$. As the beamspot is magnified by $M(140~\text{m}) = M_{\text{ZDC}}$, the projected impact point $(v_{x,\text{ZDC}},v_{y,\text{ZDC}})$ on the ZDC surface is calculated as
\begin{linenomath}
\begin{align}
v_{x,\text{ZDC}} &= M_{\text{ZDC}} \cdot (v_x-x_0) + x_0, \\
v_{y,\text{ZDC}} &= M_{\text{ZDC}} \cdot (v_y-y_0) + y_0.
\end{align}
\end{linenomath}
Finally the direction vector calculated from the impact point and the interaction point is rotated by half of the $\alpha$ crossing angle in the x-z plane. The beamspot and beam parameters are summarized in Table \ref{tab:param}.

\begin{table}[t]
\caption{Beam and beamspot parameters used in the simulation.}
\centering
\begin{tabular}{lc}
\hline
\multicolumn{1}{c}{Parameter} & \multicolumn{1}{c}{Value} \\
\hline
$\beta^*$ & 60~cm \\
$\alpha$ & 280~$\mu$rad \vspace{5px} \\

$x_0$	&	0.58~mm \\
$y_0$	&	1.05~mm \\
$z_0$	&	16~mm  \vspace{5px} \\

$\sigma_x$ & 0.013~mm \\ 
$\sigma_y$ & 0.013~mm \\
$\sigma_z$ & 47~mm \\
\hline
\end{tabular}
\label{tab:param}
\end{table}

\begin{figure}[t]
\centering
\includegraphics[width=0.32\textwidth]{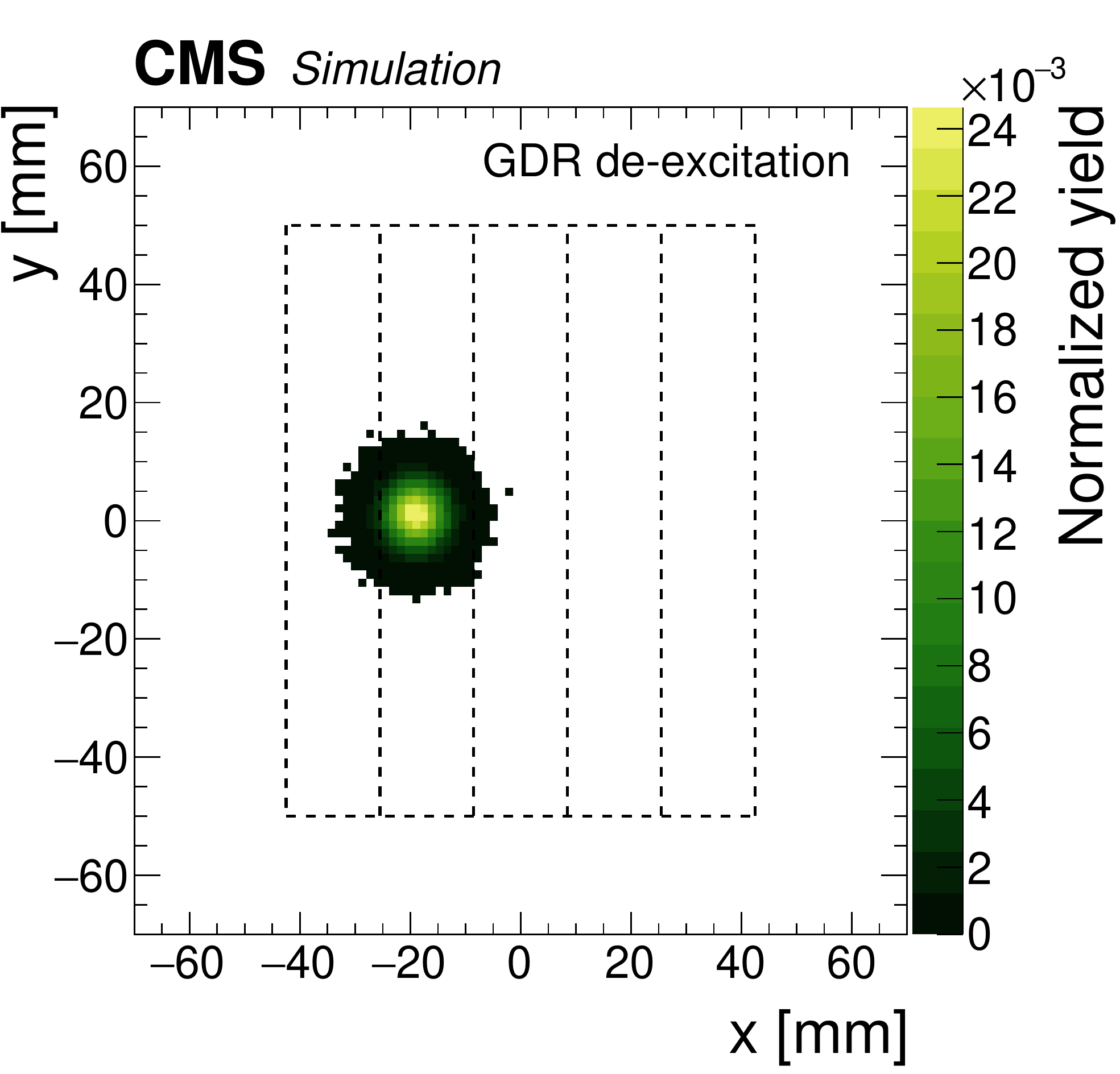}
\includegraphics[width=0.32\textwidth]{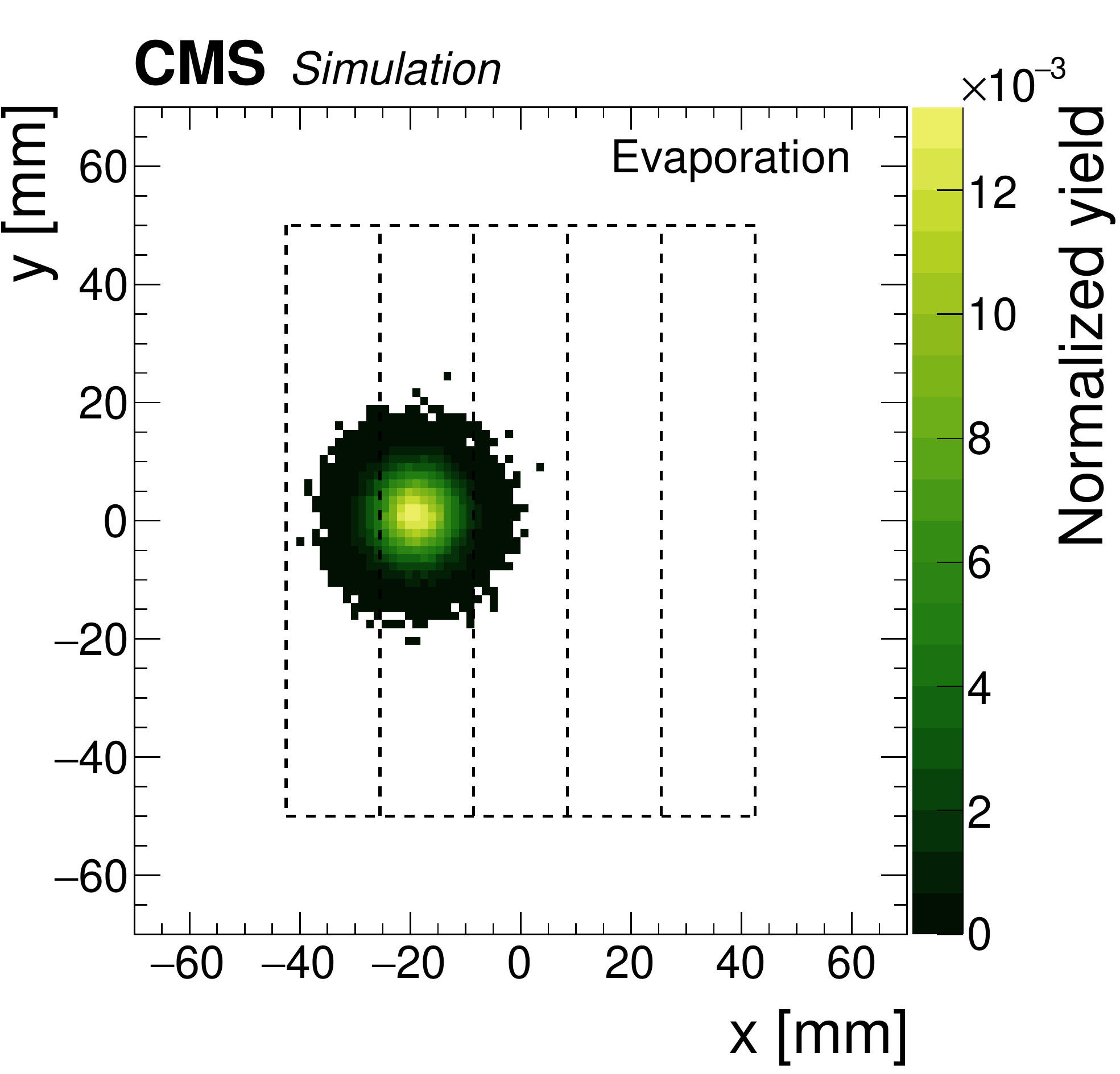}
\includegraphics[width=0.32\textwidth]{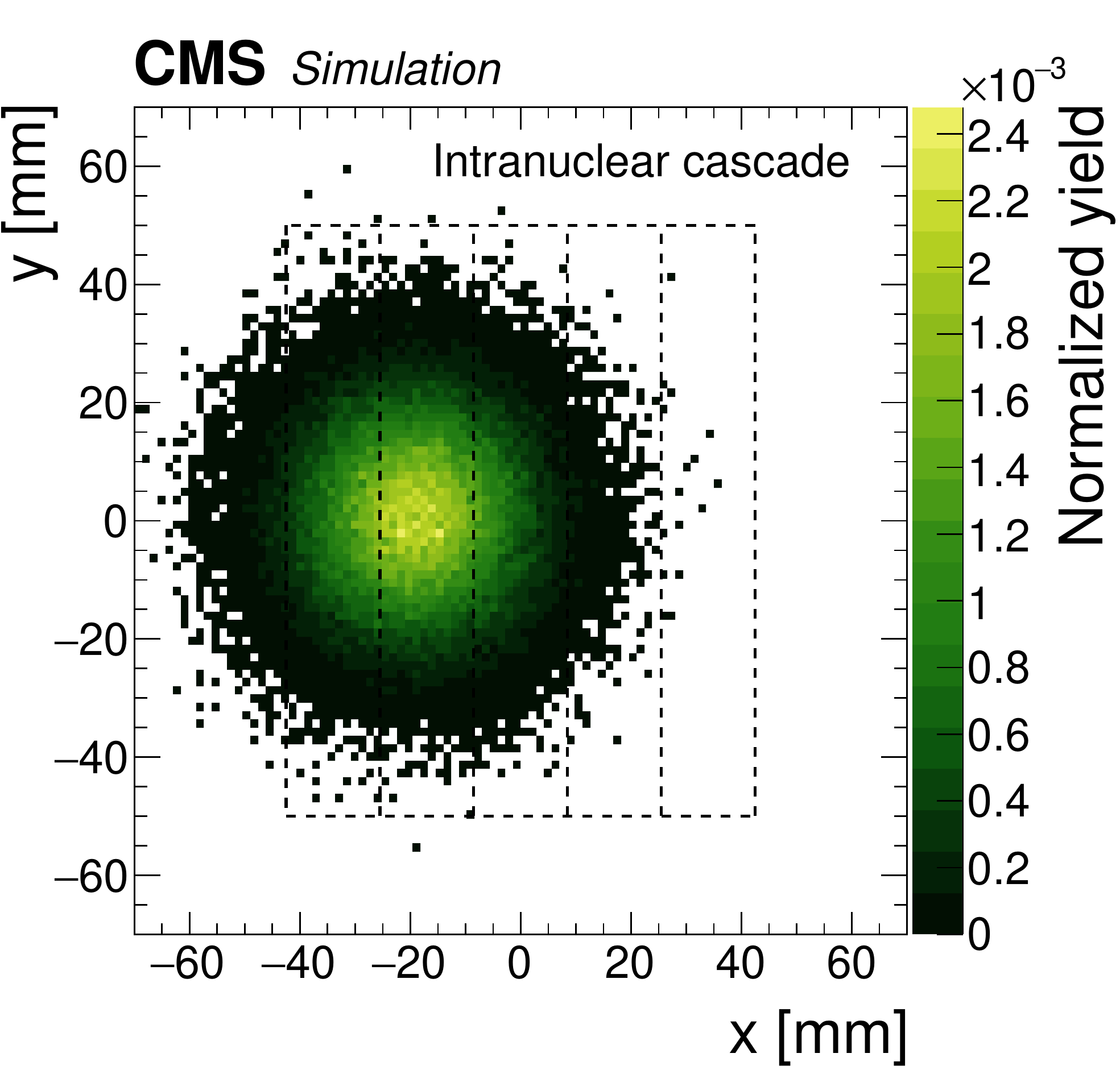}
\caption{The projected impact points on the ZDC surface of neutrons originating from GDR (left), evaporation (middle) and intranuclear cascade (right) processes.}
\label{fig:xy}
\end{figure}

\begin{figure}[t]
\centering
\includegraphics[width=0.32\textwidth]{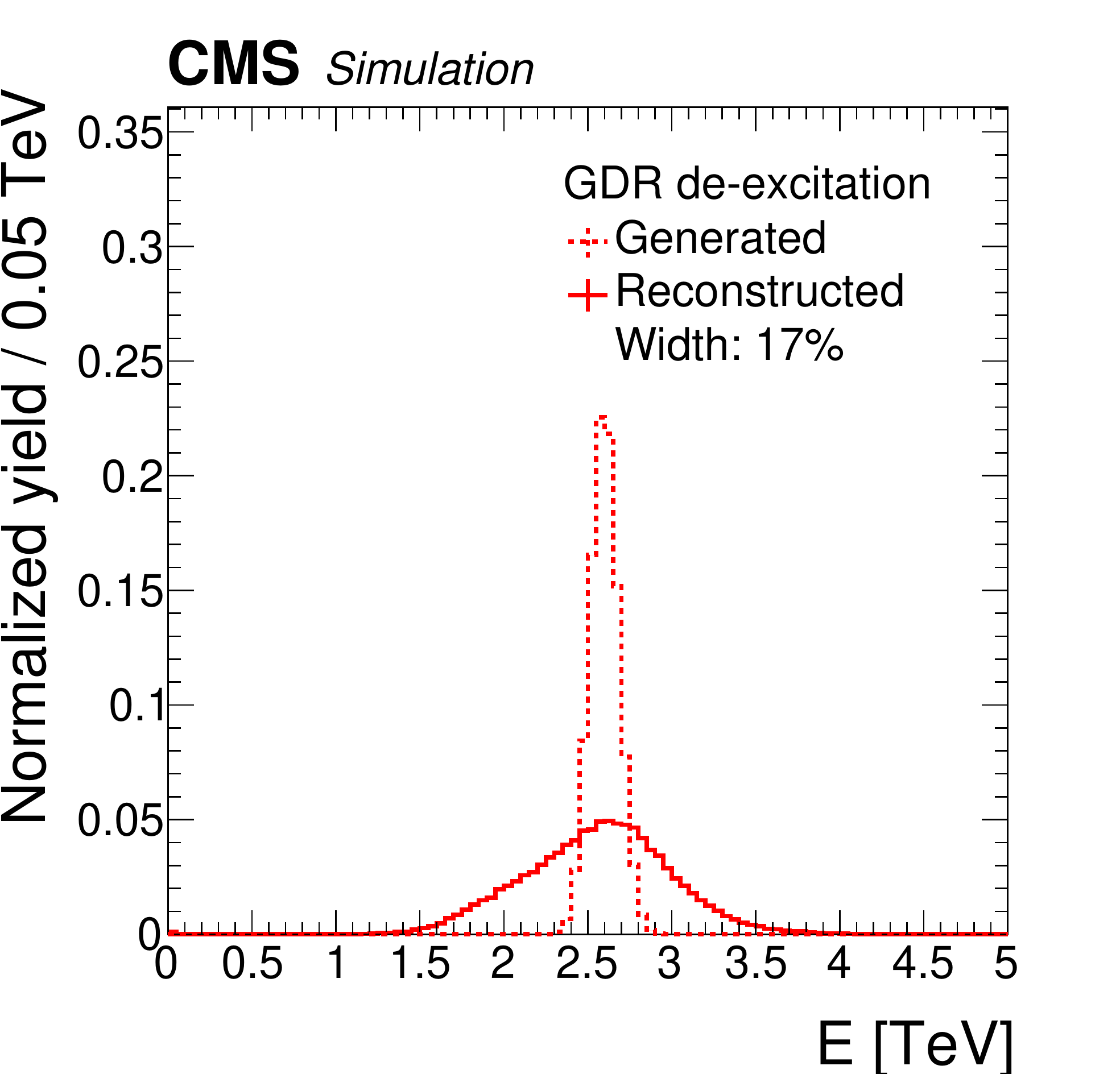}
\includegraphics[width=0.32\textwidth]{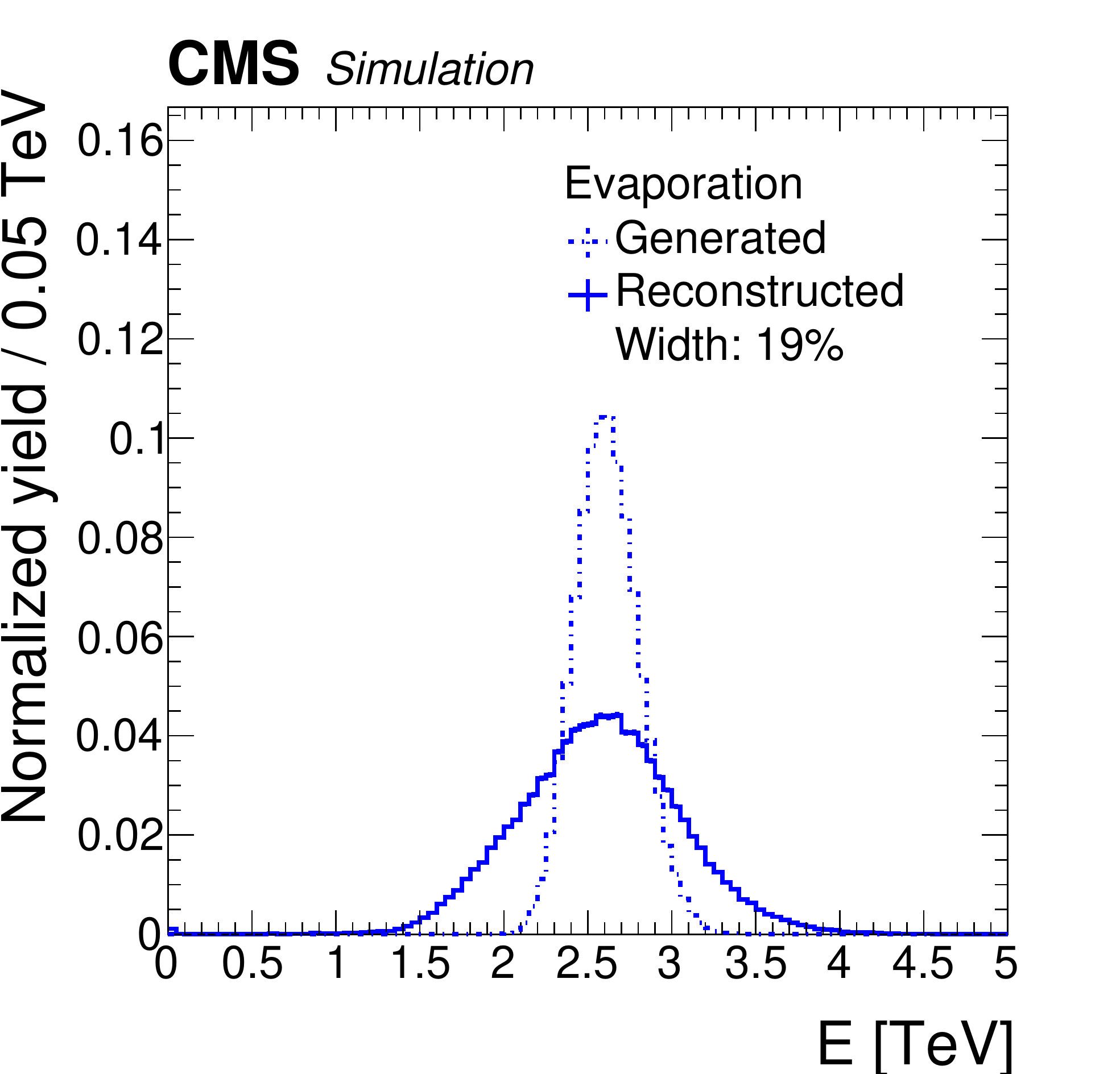}
\includegraphics[width=0.32\textwidth]{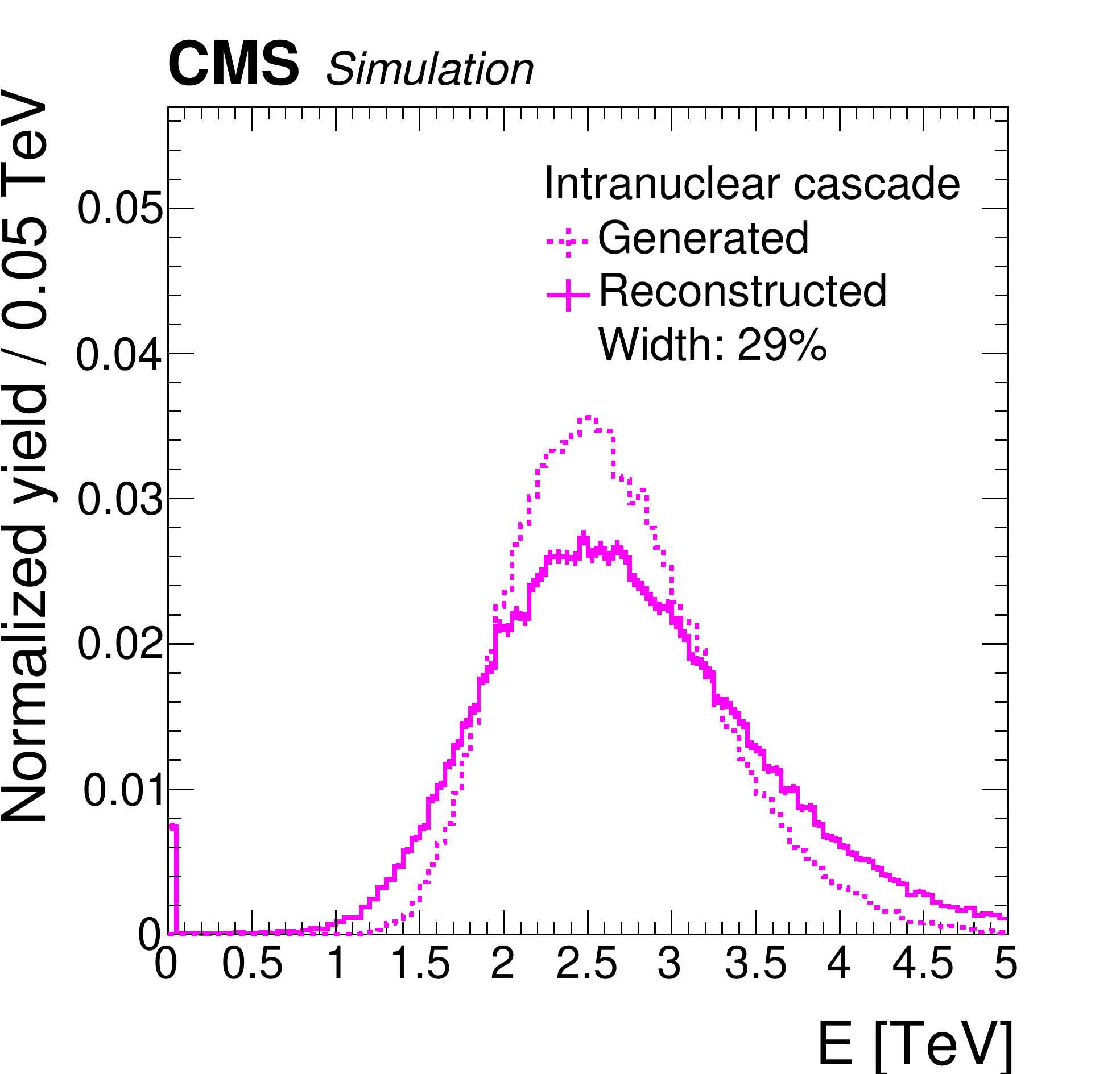}
\caption{Generator-level and reconstructed energy distributions of simulated neutrons emitted from a 2.56~TeV energy Pb ion via GDR (left), evaporation (middle) and intranuclear cascade (right) processes.}
\label{fig:sim_energy}
\end{figure}

The projected impact points for the three assumed neutron emission scenarios are shown in Fig.\ \ref{fig:xy}. The corresponding geometrical acceptance is larger than $98\%$ for all processes. The generated and observed energy distribution for three different types of very forward neutrons are summarized in Fig.\ \ref{fig:sim_energy}. It can be concluded that in the case of evaporation and GDR neutrons, the resolution is dominated by the detector response, whereas for cascade neutrons the energy spread dominates due to the large Maxwell-Boltzmann temperature.

\section{Signal extraction}

\begin{figure}[t]
\centering
\includegraphics[width=0.49\textwidth]{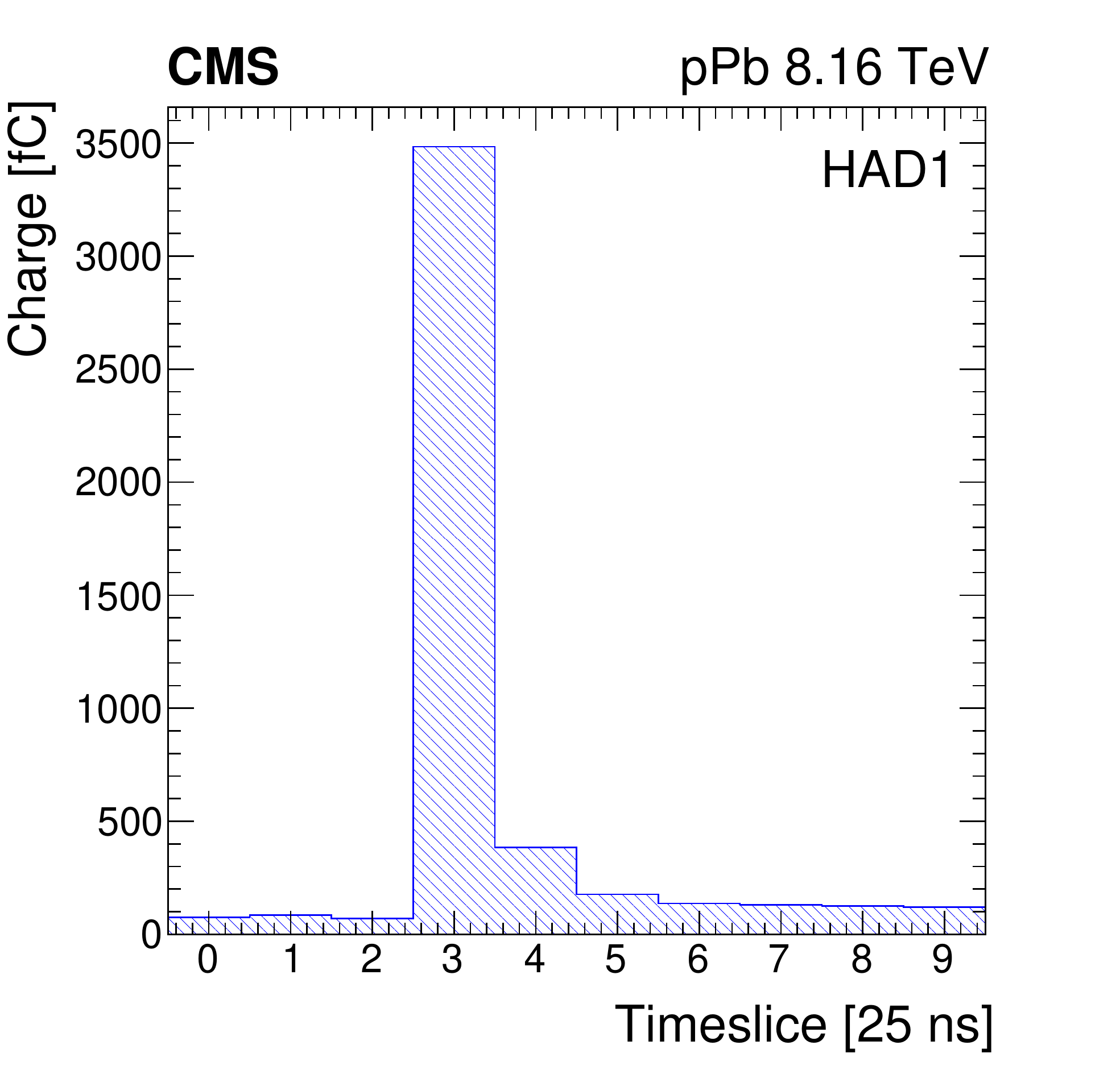}
\caption{A typical ZDC signal shape.}
\label{fig:signal}
\end{figure}

A typical signal shape in a given channel $i$ is shown in Fig.\ \ref{fig:signal}. A simple way to extract the $a_i$ signal amplitude corresponding to this shape is:
\begin{linenomath}
\begin{align}
a_i = q_i[3] - q_{\text{ped},i},
\label{eq:simple}
\end{align}
\end{linenomath}
where $q_i[t]$ is the charge value in the $t$ timeslice, and $q_{\text{ped},i}$ is the pedestal calculated as
\begin{linenomath}
\begin{align}
q_{\text{ped},i} = \frac{1}{2} \left[ q_i[0] + q_i[1] \right].
\label{eq:ped}
\end{align}
\end{linenomath}
The signals in TS0 and TS1 are used in the pedestal estimation to minimize the inclusion of the tail of the main signal. When the signal is saturated, the ZDC signal tail is calculated, defined as
\begin{linenomath}
\begin{align}
a_i^{\text{tail}} = R_i \cdot q_i[4] - q_{\text{ped},i},
\end{align}
\end{linenomath}
where the $R_i$ factors are calculated from the distributions of $(q_i[3]-q_{\text{ped},i})/(q_i[4]-q_{\text{ped},i})$ values in non-saturating signals.

\begin{figure}[t]
\centering
\includegraphics[width=0.49\textwidth]{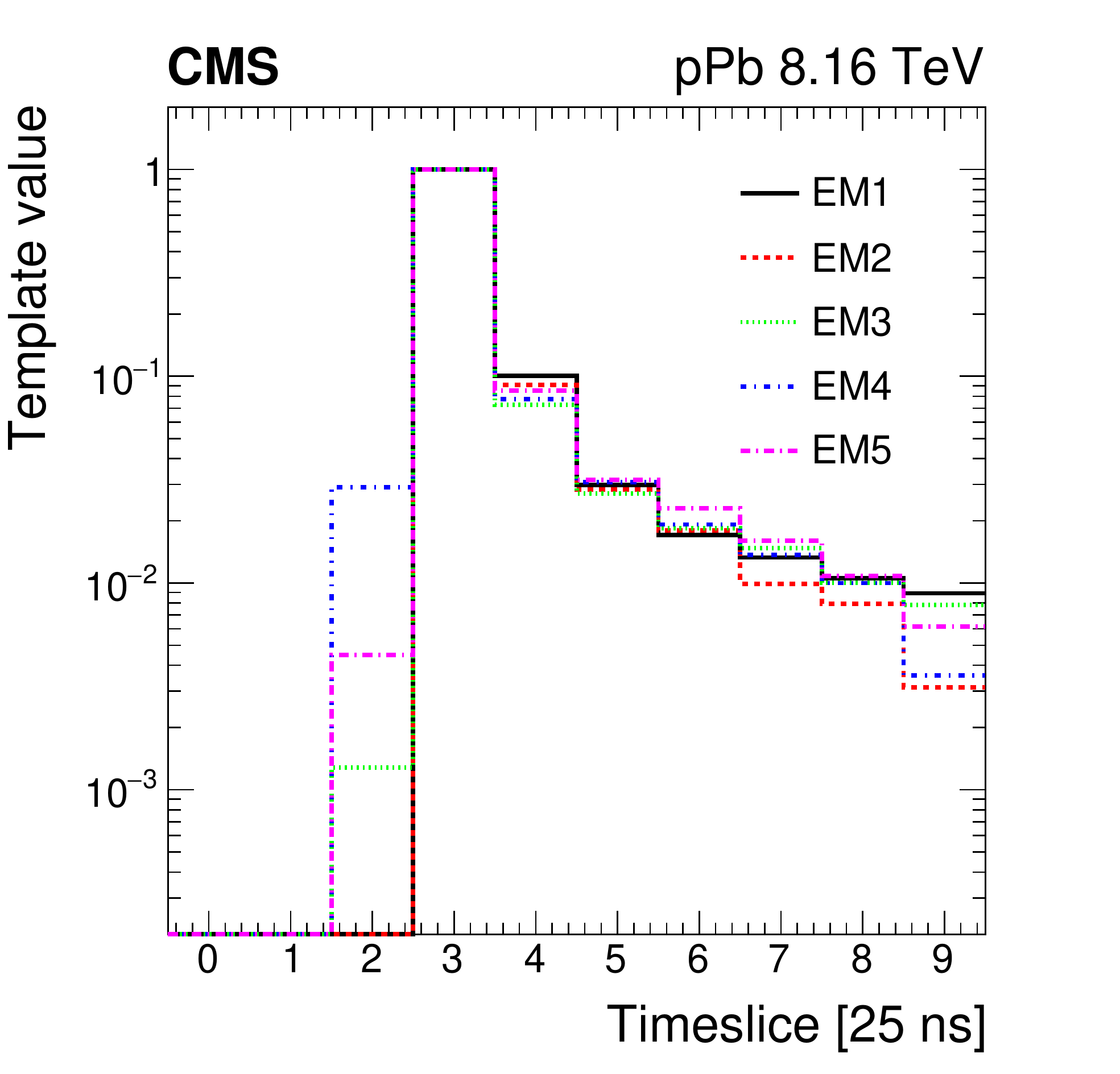}
\includegraphics[width=0.49\textwidth]{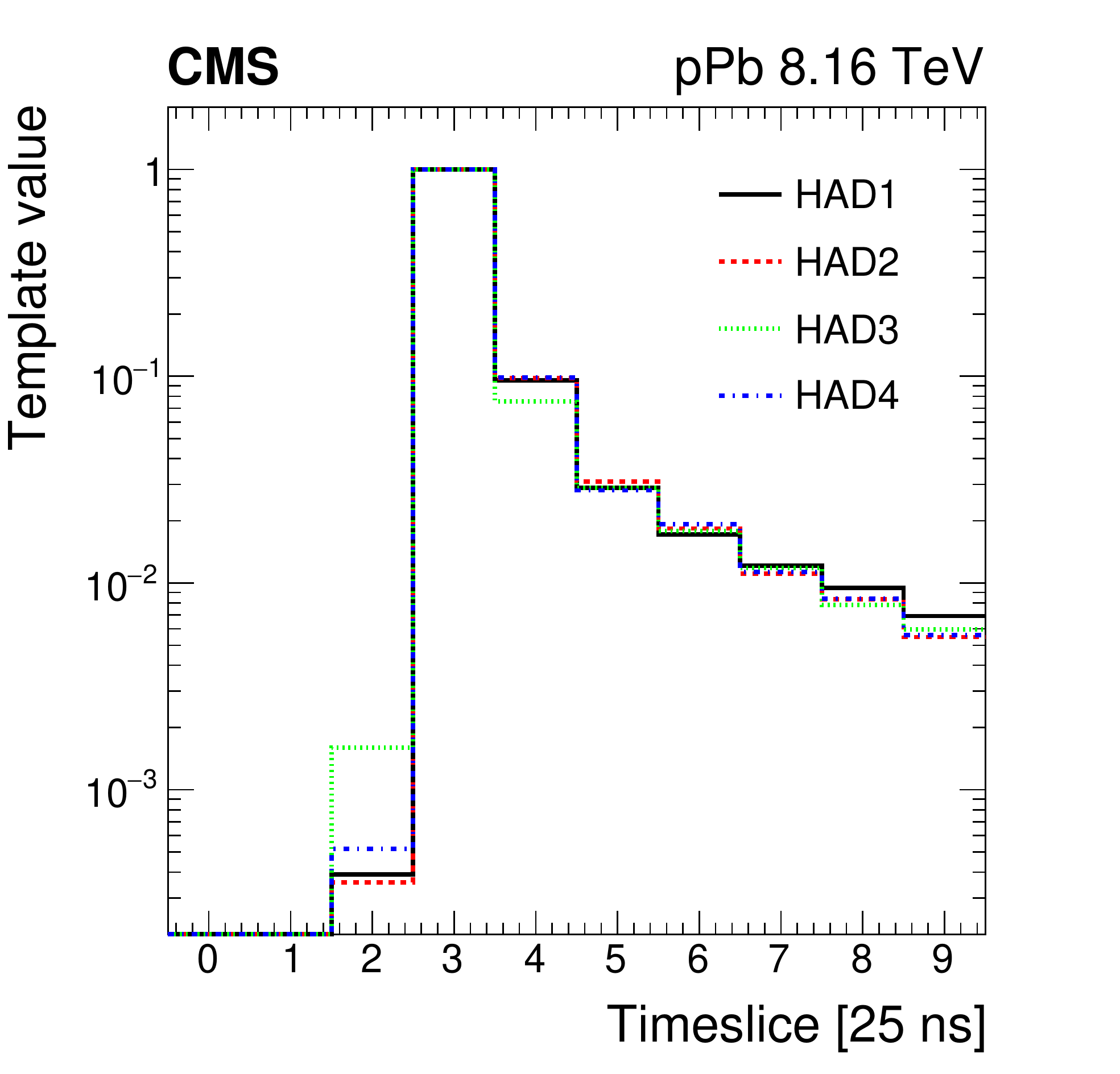}
\caption{Average signal shapes of EM channels (left) and HAD channels (right). The first six timeslices are taken as the template of the main signal, whereas the values from TS4 to TS9 are used to construct the template of the pre-pileup signals.}
\label{fig:template}
\end{figure}

Due to the 100 ns bunch spacing, further pPb collisions may occur 100~ns before or after the main signal, which are called pre-pileup and post-pileup collisions. Therefore additional signals may be present in TS7 (post-pileup) and in the timeslice preceding TS0 (pre-pileup). When a pre-pileup signal is present, Eq.\ (\ref{eq:simple}) will overestimate the pedestal value. In order to treat the events with feed-off from pre-pileup signals, a template fitting method similar to that described in Ref.\ \cite{Sirunyan:2020pmc} is used. In the following description of this method the channel indices $i$ are dropped for the sake of simplicity and vector notation is used: all vector indices correspond to a given timeslice. In order to be able to fully model the pre-pileup shape and eliminate all contribution from the post-pileup signal, only the first six timeslices are used in the fit, thus all of the following vectors are 6-dimensional. The measured signal values in a single event are denoted by $\mathbf{q}$, whereas $\mathbf{t}$ and $\mathbf{t'}$ stands for the main and the pre-pileup signal template respectively. The template for a given channel is constructed by averaging many signal shapes from which the pedestal described by Eq.\ (\ref{eq:ped}) is subtracted in each timeslice and their integral is fixed to unity in TS3. In the averaging those events are used, that have larger than $4000$~fC signal in TS3 and have no pre- or post-pileup present. The pre-pileup events are rejected by requiring TS0 and TS1 to have less than $50$~fC charge, whereas the post-pileup events are rejected by requiring charge values decreasing monotonically from TS5. The average template shapes for the different channels are shown in Fig.\ \ref{fig:template}. The first six timeslices are denoted as $\mathbf{t}$, whereas the values from TS4 to TS9 are used to construct $\mathbf{t'}$.

\begin{figure}[t]
\centering
\includegraphics[width=0.32\linewidth]{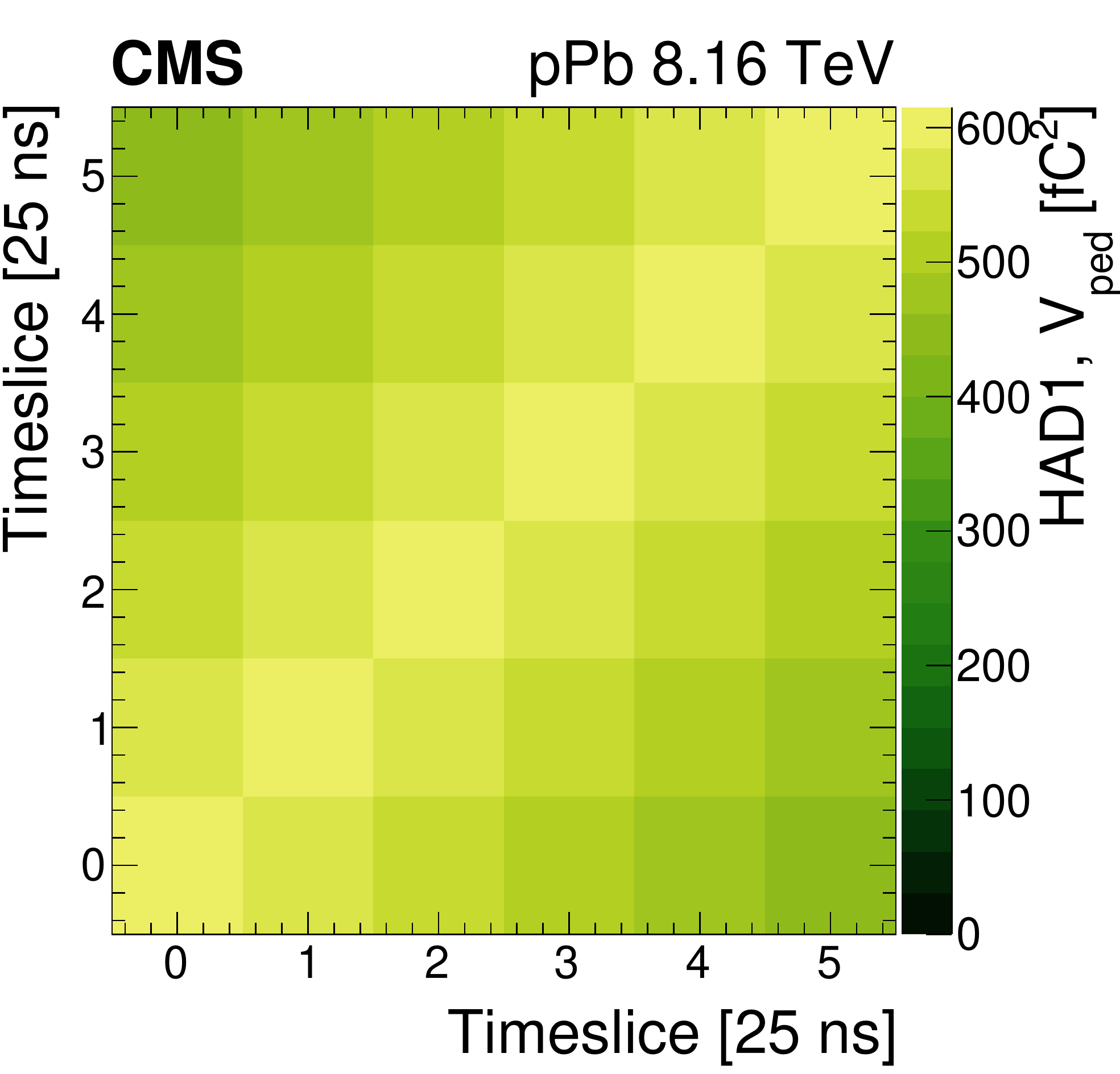}
\includegraphics[width=0.32\linewidth]{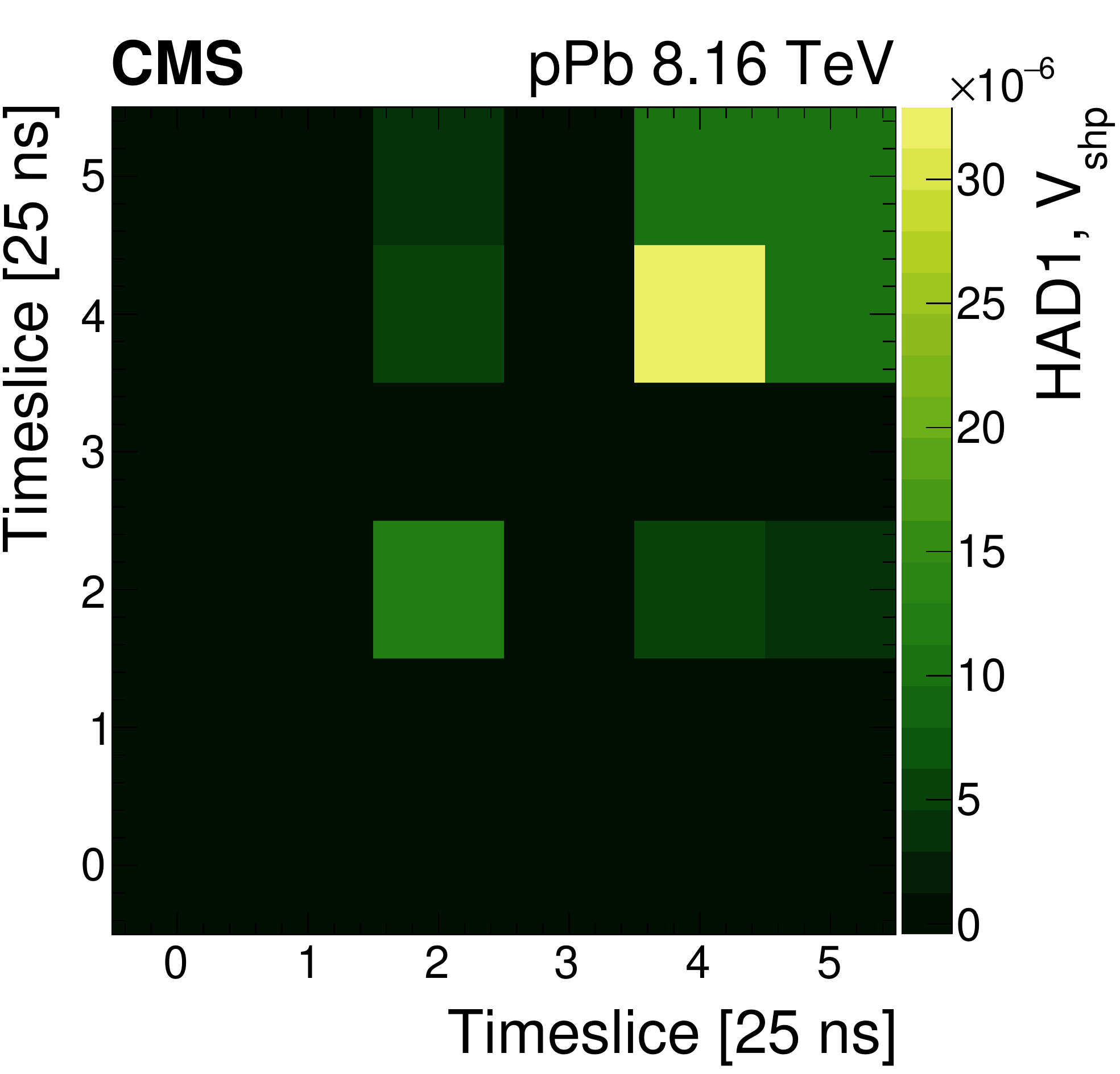}
\includegraphics[width=0.32\linewidth]{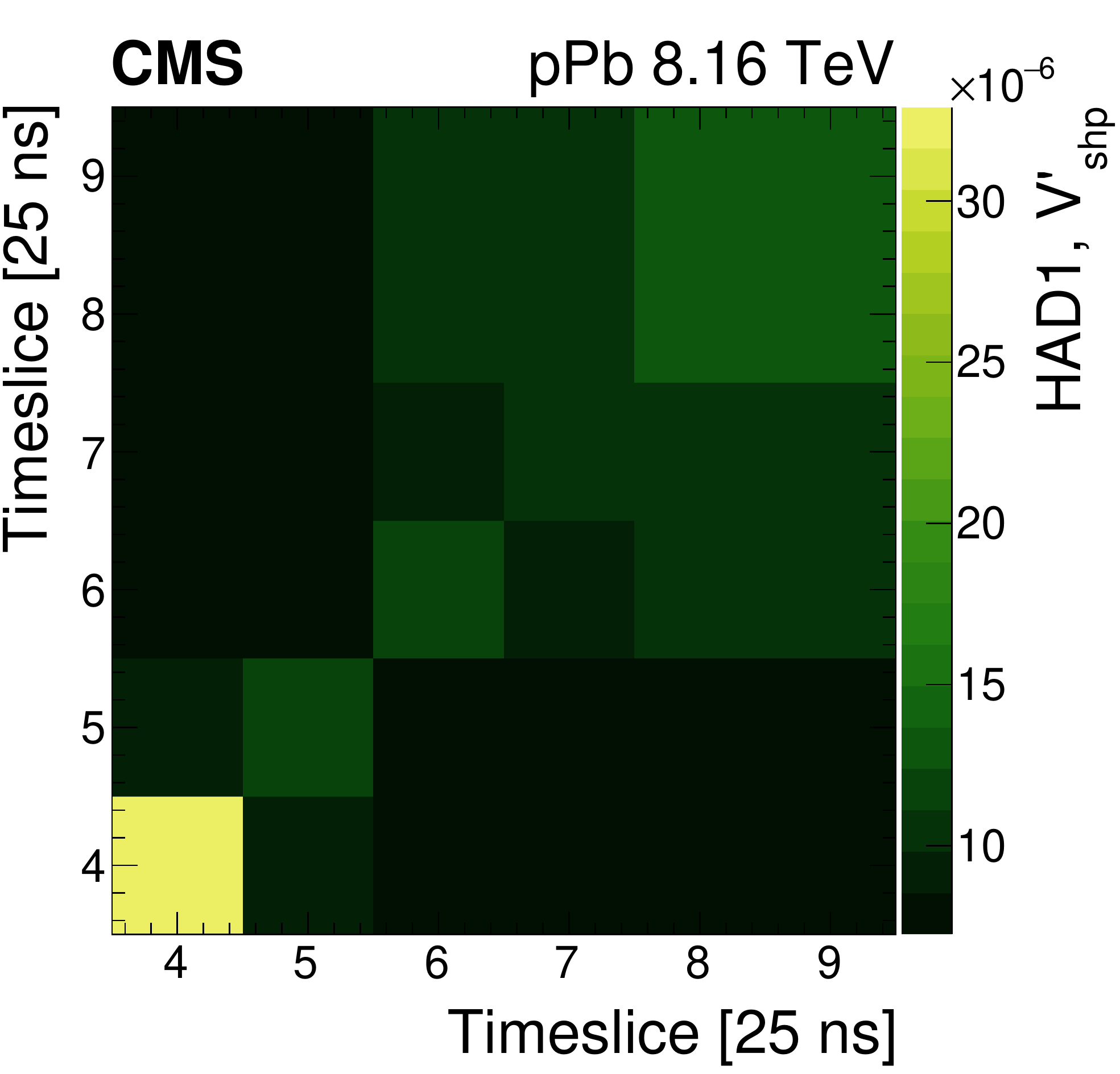}
\caption{Covariance matrices of pedestal fluctuations (left), in-time (middle) and pre-pileup pulse shape (right) of HAD1 channel. The covariance matrices of other channels look similar. The unit of the elements of pedestal covariance matrix is fC$^2$, whereas the pulse shape matrices do not have a unit as they are calculated from normalized signal shapes.}
\label{fig:cov}
\end{figure}

The amplitude of the signal is then calculated by minimizing the following $\chi^2$-like expression:
\begin{linenomath}
\begin{align}
\chi^2 = (\mathbf{q}-a\mathbf{t}-b\mathbf{t'} - c\mathbf{1})^{\mathsf{T}} \mathbf{V}^{-1} (\mathbf{q}-a\mathbf{t}-b\mathbf{t'}-c\mathbf{1}),
\label{eq:chi2}
\end{align}
\end{linenomath}
where $\mathbf{V}$ is the covariance matrix, $\mathbf{1}$ is a 6-element vector with all components equal to $1$, $a$ is the main signal amplitude, $b$ is the amplitude of the pre-pileup signal and $c$ is the pedestal. This minimization is a fit, with $a$, $b$, and $c$ as free parameters. There are three contributions to $\mathbf{V}$: (i) the digitization uncertainty of the measured signal, (ii) the fluctuations of the pedestal, where the off-diagonal elements should also be considered, and (iii) the uncertainty of template shapes due to digitization and the uncertainty in the timing of the signals.
The term corresponding to the digitization uncertainty is approximated as:
\begin{linenomath}
\begin{align}
V_{\text{dig},ij} = \delta_{ij} \cdot \frac{\Delta q_i^2}{12},
\end{align}
\end{linenomath}
where $\Delta q_i$ is the width of the charge range corresponding to the measured digital value provided by the analog-digital converter in timeslice $i$ and $\delta_{ij}$ denotes the Kronecker delta. The fluctuations of the pedestal originates from the dark current of the PMTs, the pickup noise of the cables, and the thermal noise of the cables depending on the temperature and the capacitance of the cables. The pedestal is typically around 30-70~fC, whereas in single neutron events the amplitudes of the signals in TS3 are in the 100-300~fC range. In multineutron events, signal values are generally higher, and as a consequence the effect of the noise is less significant. The covariance term of pedestal fluctuations is calculated from non-collision events using the sample covariance formula:
\begin{linenomath}
\begin{align}
V_{\text{ped},ij} \approx \frac{\sum_{k=1}^N (q_i^k - \bar{q})(q_j^k - \bar{q})}{N-1},
\end{align}
\end{linenomath}
where $q_i^k$ is the signal value in the timeslice $i$ in the event $k$, $\bar{q}$ is the average pedestal level, and $N$ is the total number of events. An example for a $\mathbf{V}_{\text{ped}}$ matrix is shown in the left panel of Fig.\ \ref{fig:cov}. The large off-diagonal elements indicate a low frequency variation of the pedestal.
The pulse shape covariance matrices $\mathbf{V}_{\text{shp}}$ and $\mathbf{V}'_{\text{shp}}$, corresponding to the main and pre-pileup signal respectively, are calculated similarly using the collision events that were used for the determination of the template, and are shown in the middle and right panel of Fig.\ \ref{fig:cov} respectively. The final covariance matrix is defined as
\begin{linenomath}
\begin{align}
\mathbf{V} = \mathbf{V}_{\text{dig}} + \mathbf{V}_{\text{ped}} + a^2 \, \mathbf{V}_{\text{shp}} + b^2 \, \mathbf{V}'_{\text{shp}}.
\end{align}
\end{linenomath}
Since the parameters $a^2$ and $b^2$ introduce a fourth order term in the $\chi^2$ expression, they are estimated as $a \approx q[3]$ and $b \approx R_i q[0]$, therefore they do not spoil the linearity of the equations derived below.

\begin{figure}[t]
\centering
\includegraphics[width=0.49\textwidth]{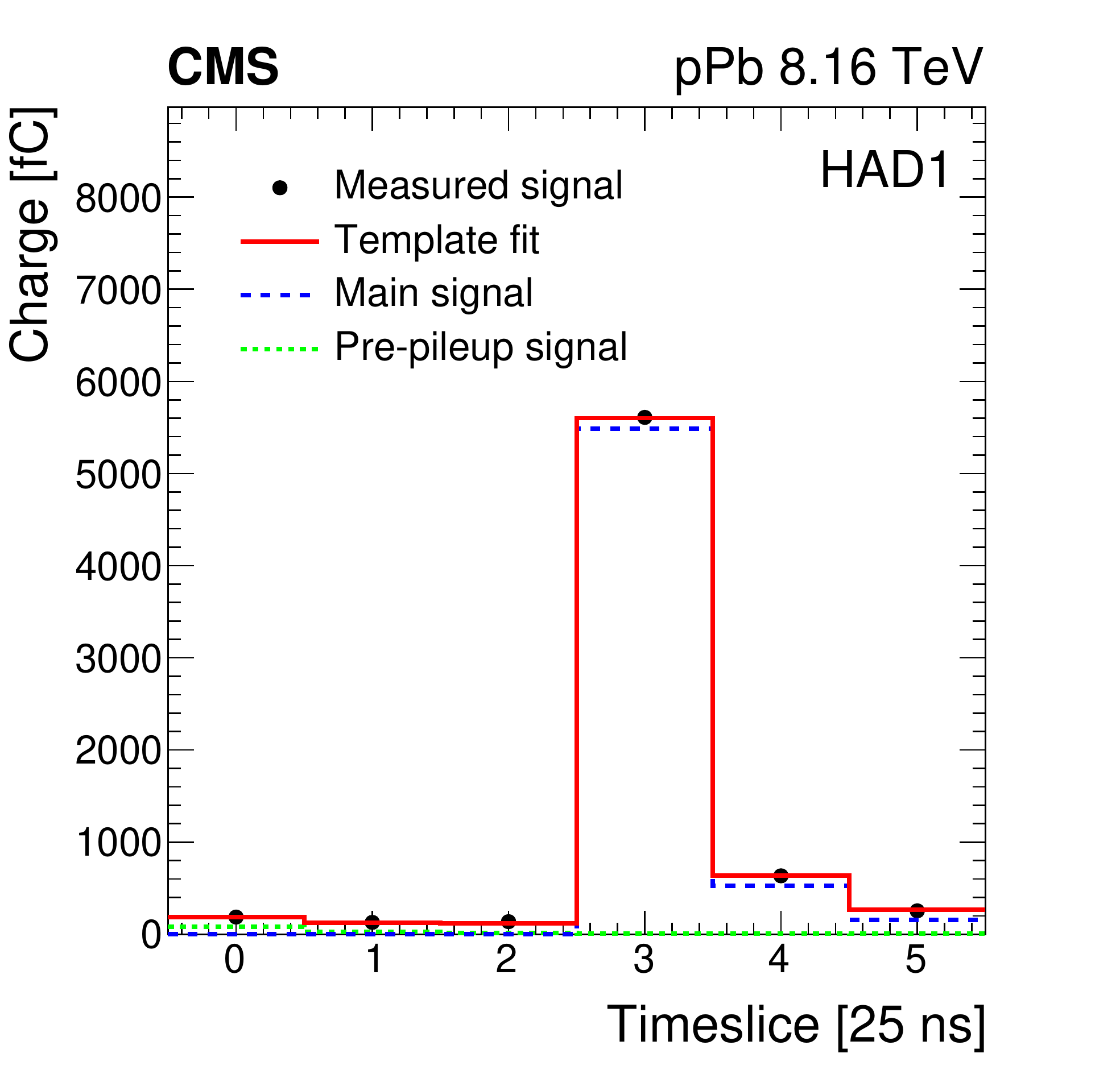}
\includegraphics[width=0.49\textwidth]{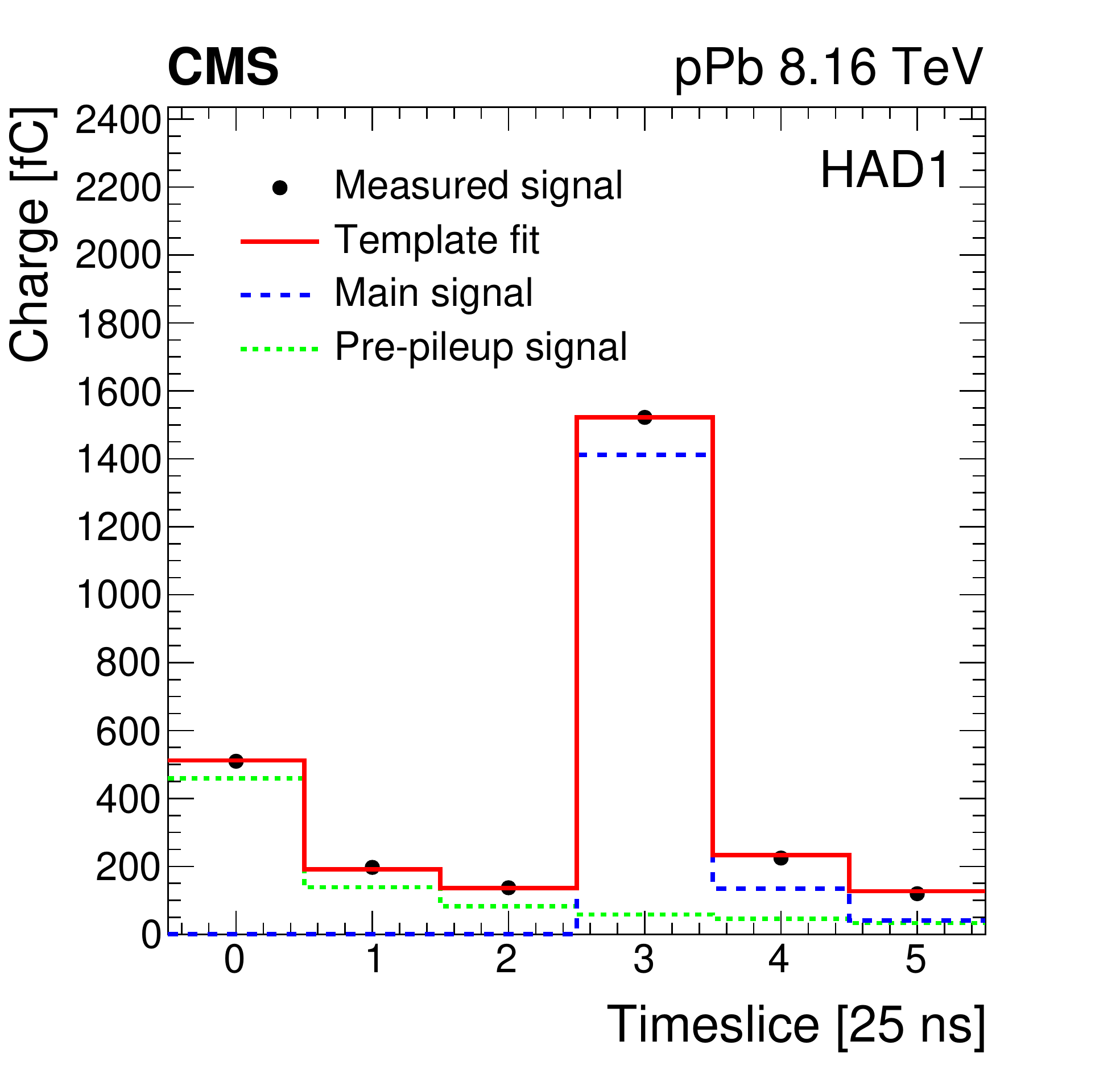}
\caption{Examples of template fits without a pre-pileup signal (left) and with a pre-pileup signal occurring $100$~ns before the main signal (right).}
\label{fig:example}
\end{figure}

The optimal parameter values can be calculated by taking the partial derivatives of this expression with respect to the parameters:
\begin{linenomath}
\begin{align}
0 &= \frac{\mathrm{d}\chi^2}{\mathrm{d}a} = -2\mathbf{t} \mathbf{V}^{-1} (\mathbf{q}-a\mathbf{t}-b\mathbf{t}'-c\mathbf{1}), \\
0 &= \frac{\mathrm{d}\chi^2}{\mathrm{d}b} = -2\mathbf{t}' \mathbf{V}^{-1} (\mathbf{q}-a\mathbf{t}-b\mathbf{t}'-c\mathbf{1}), \\
0 &= \frac{\mathrm{d}\chi^2}{\mathrm{d}c} = -2\mathbf{1} \mathbf{V}^{-1} (\mathbf{q}-a\mathbf{t}-b\mathbf{t}'-c\mathbf{1}).
\end{align}
\end{linenomath}
Now $\mathbf{A}$, $\mathbf{v}$ and $\mathbf{x}$ are defined as
\begin{linenomath}
\begin{align}
\mathbf{A} = \left[ \begin{matrix}
 \mathbf{t}^{\mathsf{T}} \mathbf{V}^{-1} \mathbf{t} &&& \mathbf{t}'^{\mathsf{T}} \mathbf{V}^{-1} \mathbf{t} &&& \mathbf{1}^{\mathsf{T}} \mathbf{V}^{-1} \mathbf{t} \\
 \mathbf{t}^{\mathsf{T}} \mathbf{V}^{-1} \mathbf{t}' &&& \mathbf{t}'^{\mathsf{T}} \mathbf{V}^{-1} \mathbf{t}' &&&  \mathbf{1}^{\mathsf{T}} \mathbf{V}^{-1} \mathbf{t}' \\
 \mathbf{t^{\mathsf{T}}} \mathbf{V}^{-1} \mathbf{1} &&& \mathbf{t}'^{\mathsf{T}} \mathbf{V}^{-1} \mathbf{1} &&& \mathbf{1}^{\mathsf{T}} \mathbf{V}^{-1} \mathbf{1}
\end{matrix} \right],
\quad
\mathbf{v} = \left[ \begin{matrix}
\mathbf{q^{\mathsf{T}}} \mathbf{V}^{-1} \mathbf{t} \\
\mathbf{q^{\mathsf{T}}} \mathbf{V}^{-1} \mathbf{t}' \\
\mathbf{q^{\mathsf{T}}} \mathbf{V}^{-1} \mathbf{1}
\end{matrix}\right],
\quad
\mathbf{x} = \left[ \begin{matrix}
a \\
b \\
c
\end{matrix}\right].
\end{align}
\end{linenomath}
and the optimal parameters can be calculated by solving the
\begin{linenomath}
\begin{align}
\mathbf{A} \mathbf{x} = \mathbf{v}
\end{align}
\end{linenomath}
linear equation. Two example fit results are shown in Fig.\ \ref{fig:example}.

This method can be generalized for signals saturating in TS3 by omitting the elements from all vectors and matrices corresponding to TS3 and adding a penalty term to (\ref{eq:chi2}). Let $\mathbf{\hat{q}}$, $\mathbf{\hat{t}}$, and $\mathbf{\hat{t}}'$ be the vector of signal and template values in each timeslice, except TS3 -- thus they are 5-dimensional vectors. Then the $\hat{\chi}^2$ expression to minimize is
\begin{linenomath}
\begin{align}
\hat{\chi}^2 = (\mathbf{\hat{q}}-a\mathbf{\hat{t}}-b\mathbf{\hat{t}}' - c\mathbf{1})^{\mathsf{T}} \mathbf{\hat{V}^{-1}} (\mathbf{\hat{q}}-a\mathbf{\hat{t}}-b\mathbf{\hat{t}}' - c\mathbf{1}) + \chi^2_{\text{sat}},
\end{align}
\end{linenomath}
where, using the $q_s$ saturation value, the penalty term is:
\begin{linenomath}
\begin{align}
\chi^2_{\text{sat}} =
\begin{cases}
-2\log\left[1-\mathrm{erf}\left(\frac{q_s-a \cdot t[3] - b \cdot t'[3] -c}{\sqrt{2V_{33}}} \right)\right], & \text{if} \,\, a \cdot t[3] + b \cdot t'[3] + c < q_s,  \\
0, & \text{if} \,\, a \cdot t[3] + b \cdot t'[3] + c \geq q_s.
\end{cases}
\end{align}
\end{linenomath}
This term is introduced to penalize those fit functions that predict a smaller value than $q_s$ in TS3. As a result, the fit function will be closer to $q_s$. The first part of $\chi^2_{\text{sat}}$ is approximated by a second order polynomial, calculated using the Maclaurin expansions of $\log(1-x)$ and $\mathrm{erf}(x)$, therefore the minimization of $\hat{\chi}^2$ can also be carried similarly as the minimization of (\ref{eq:chi2}), by solving a linear equation.

\section{Calibration}
There are response differences between the individual ZDC channels because of high voltage setting, photocathode damage of the PMTs, and radiation damage. The charge of every measured channel $i$ is multiplied by a $w_i$ factor to match the different gains of the individual channels. Thus, the total energy deposited in a ZDC detector $E$ is calculated as:
\begin{linenomath}
\begin{align}
E = \sum_i w_i a_i. 
\end{align}
\end{linenomath}

\begin{figure}[t]
\centering
\includegraphics[width=0.49\textwidth]{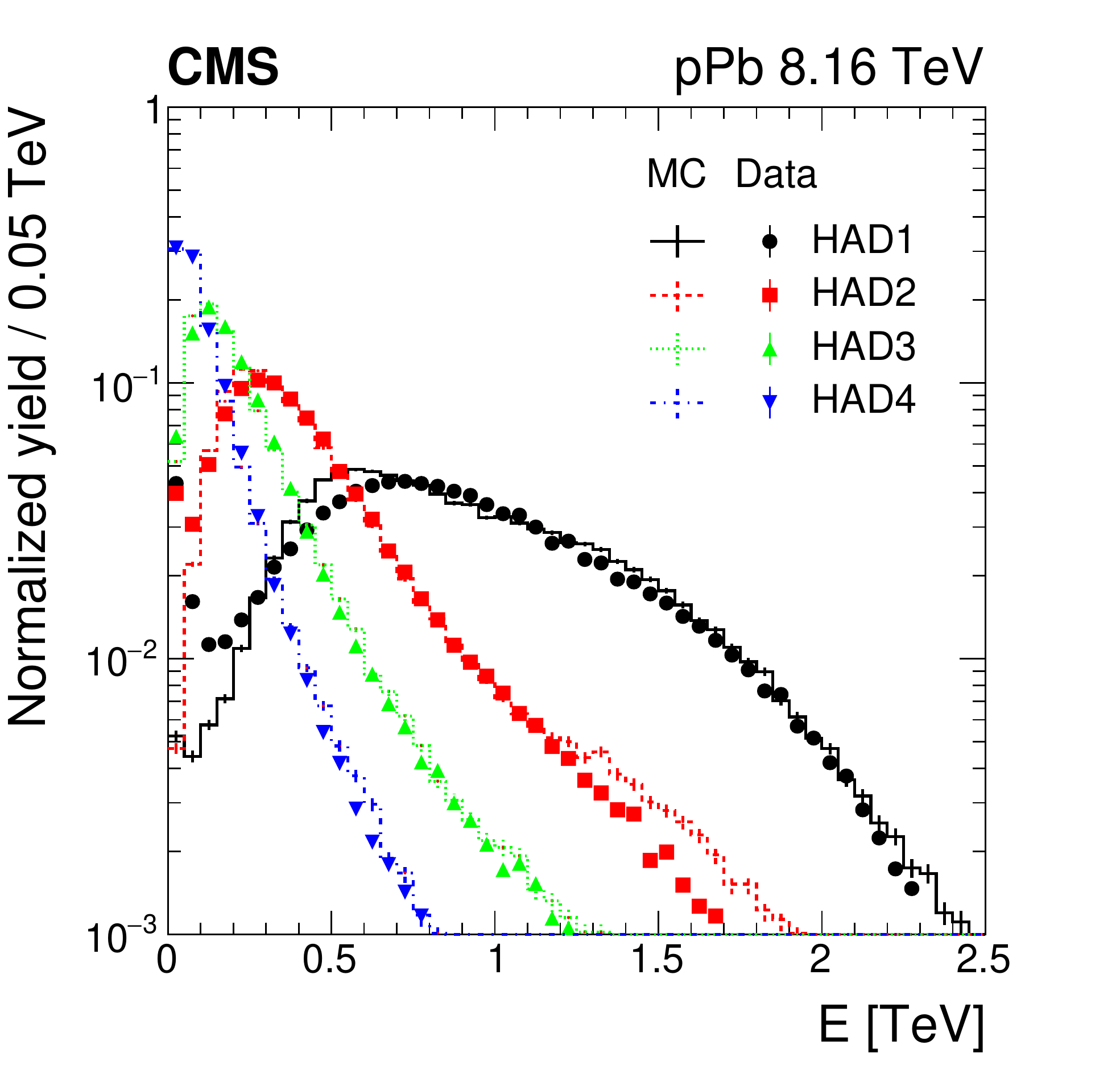}
\caption{The distribution of deposited energy in HAD section channels in simulation (lines) and data (dots).}
\label{fig:mc_ratio}
\end{figure}

First, the whole EM section is scaled to minimize the single neutron resolution, as described in Section \ref{sec:mc}. Then the gain matching constants for the hadron section channels are calculated from the comparison of detector level and simulated per-channel energy distributions using 240\,000 single neutron candidate events, as illustrated in Fig.\ \ref{fig:mc_ratio}. After the gain matching of the HAD section channels, the EM section is weighted again to match the newly calibrated HAD section. Then the HAD channel weights are refined using a more pure sample of single neutron events. Finally the EM section weights are adjusted individually by minimizing the single neutron resolution.

\begin{figure}[t]
\centering
\includegraphics[width=0.49\textwidth]{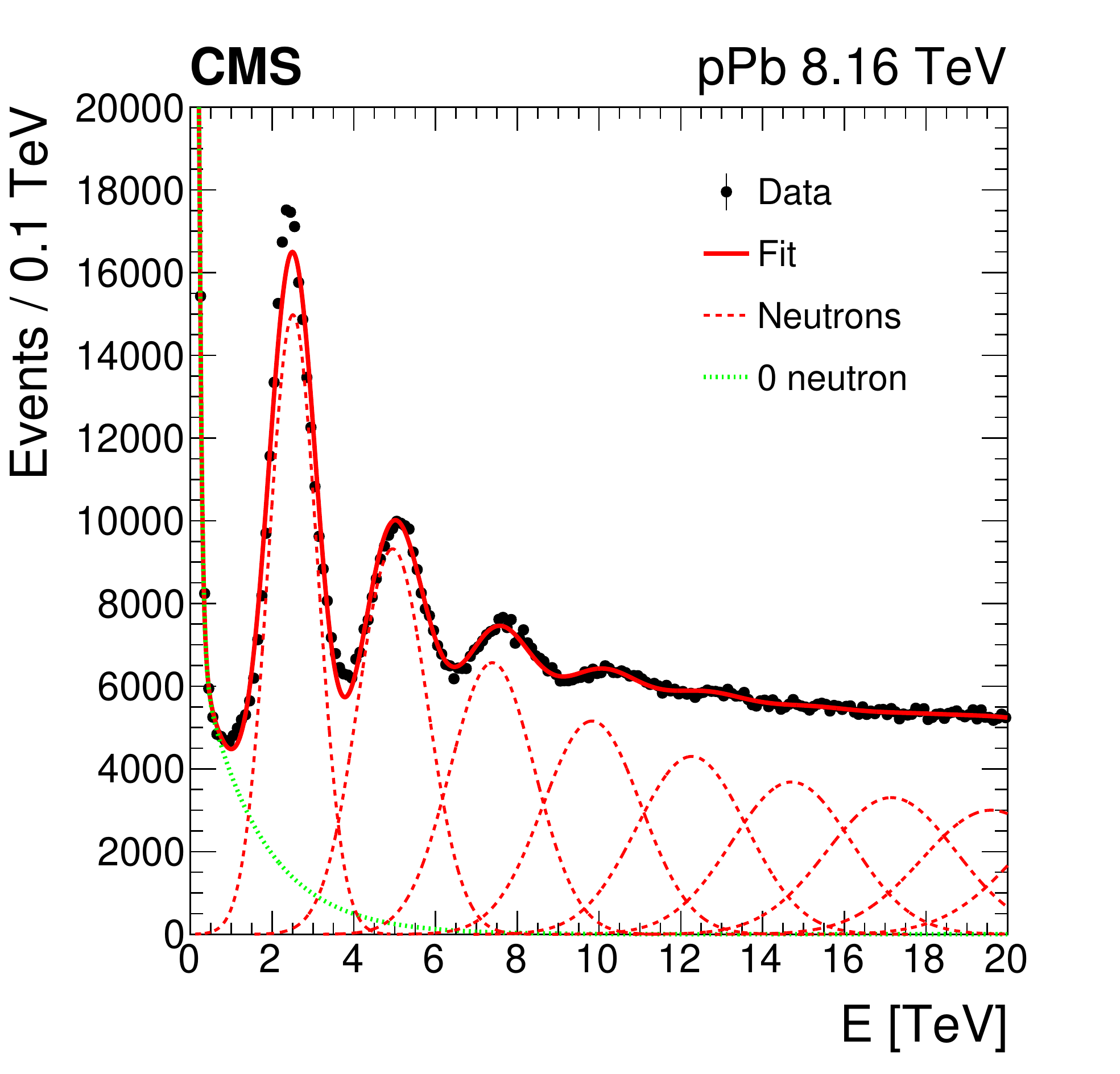}
\caption{The measured ZDC energy distribution. The three prominent peaks corresponds to single, double and triple neutron events.}
\label{fig:spectrum}
\end{figure}

The distribution of calibrated ZDC energies is shown in Fig.\ \ref{fig:spectrum}. The three prominent peaks corresponds to single, double and triple neutron events. The reason for this quasi-discrete spectrum is that the neutrons emitted from Pb ions are approximately monoenergetic due to the small Maxwell-Boltzmann temperature of neutrons and the large Lorentz boost of the Pb ion.

Assuming that the response of a single neutron can be described by a Gaussian distribution, the neutron energies are added up independently, and the zero neutron contribution is described by the sum of two exponential functions, the low-energy part of the spectrum is fitted with the sum of Gaussian distributions and two exponential distributions, describing the noise peak and the contribution of photons:
\begin{linenomath}
\begin{align}
f(E) &= a_1 \mathrm{e}^{-\lambda_1 E} + a_2 \mathrm{e}^{-\lambda_2 E} + \sum_{n=1}^{n_{\text{max}}} A_n \frac{1}{\sqrt{2\pi} \sigma_n} \mathrm{e}^{-\frac{(E-\mu_n)^2}{2\sigma_n^2}}, \\
\mu_n &= n \mu_0 + \nu, \\
\sigma_n^2 &= n\sigma_0^2,
\end{align}
\end{linenomath}
where $a_{1,2}$ and $\lambda_{1,2}$ are the parameters of the exponential functions corresponding to the zero neutron distribution, $A_n$ is the amplitude of the $n$-neutron peak, $\mu_0$, $\nu$, and $\sigma_0$ are parameters describing the positions and widths of the neutron peaks, and $n_{\text{max}}$ is the maximum number of neutrons. The fit shown in Fig.\ \ref{fig:spectrum} is performed with $n_\text{max} = 9$. The relative width of the single neutron peak at 2.56~TeV calculated from the fit is approximately $23.8\%$. This includes both the detector resolution and the additional widening from physics processes as demonstrated in Fig.\ \ref{fig:sim_energy}.

\begin{figure}[t]
\centering
\includegraphics[width=0.49\textwidth]{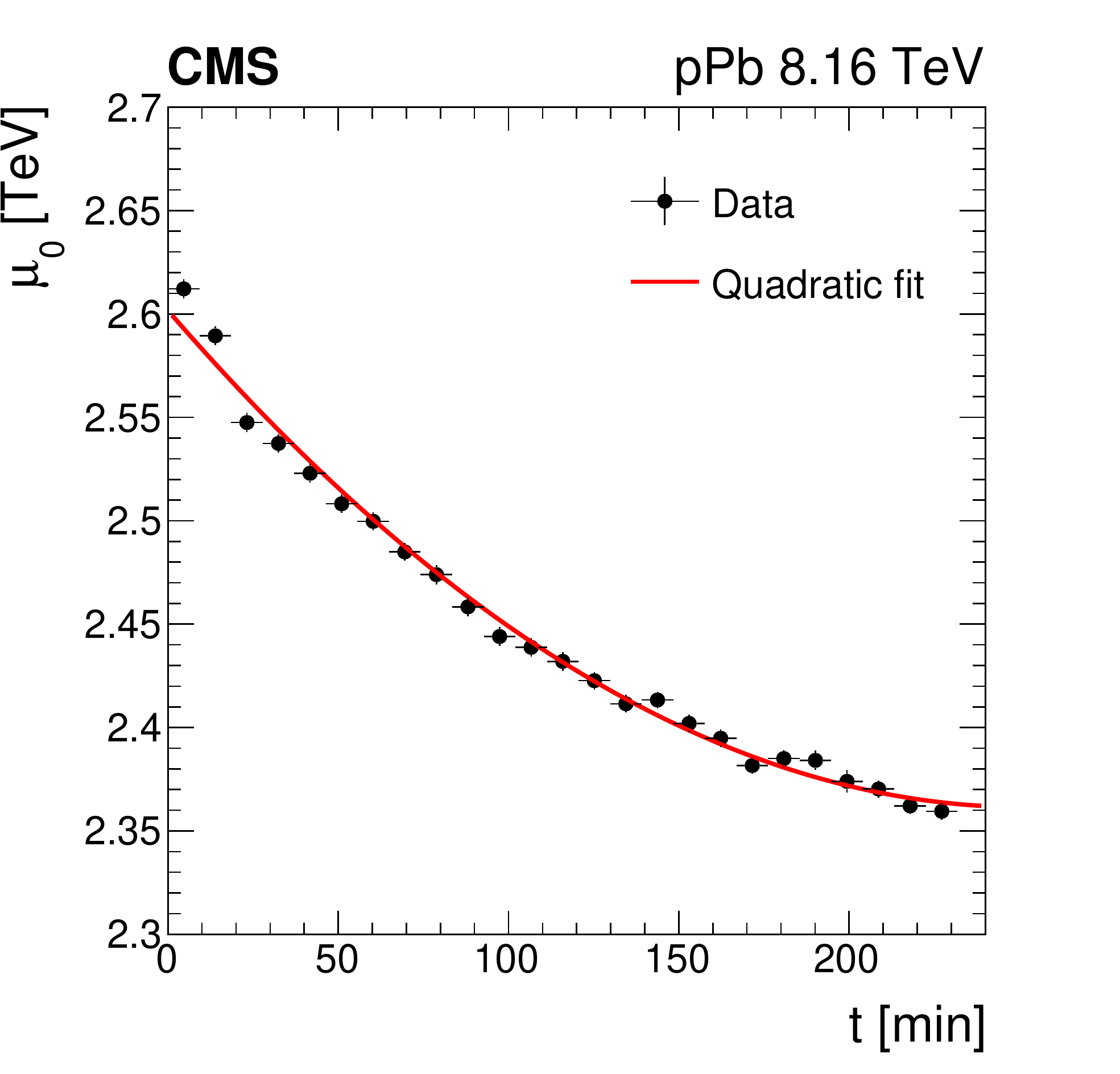}
\includegraphics[width=0.49\textwidth]{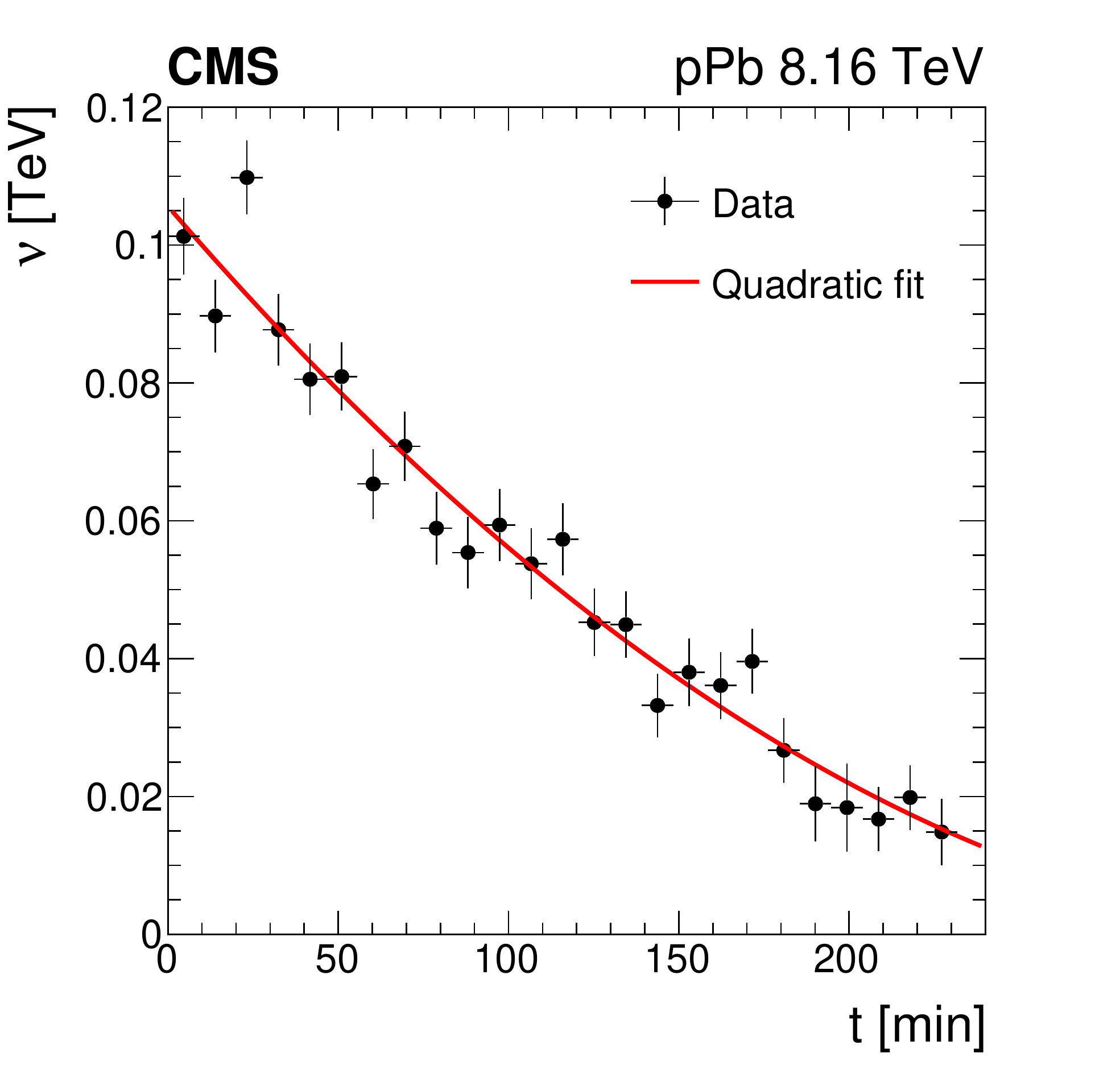}
\caption{The dependence of $\mu_0$ and $\nu$ parameters on time since the beginning of the run. The time-dependence of both $\mu_0$ and $\nu$ is fitted with a second order polynomial.}
\label{fig:lumi}
\end{figure}

Finally the time-dependence of the $\mu_0$ and $\nu$ parameters is studied. It is found that these quantities depend on time in the given run as shown in Fig.\ \ref{fig:lumi}. The reasons for this are effects that depend on the instantaneous luminosity, like the activation of the detector and the degradation of the beam quality. Second order polynomials $\mu_0(t)$ and $\nu(t)$ are fitted to describe the time-dependence. Using these, the  measured total ZDC energy can be corrected on an event-by-event basis:
\begin{linenomath}
\begin{align}
E_{\text{corr}} = \frac{E - \nu(t)}{\mu_0(t)} \times 2.56~\text{TeV}.
\end{align}
\end{linenomath}
After applying this correction, the relative width of the single neutron peak is reduced to $23.4\%$.

\section{Correction for in-time pileup}
Simultaneous pPb collisions (in-time pileup) shift the ZDC energy spectrum to higher values, which causes a rise in the tail of the ZDC energy distribution. In this paper a deconvolution method is applied to remove these multicollision events from the final distribution. A similar method was used in \cite{Laszlo:2016rll}. The probability of having $k$ number of interactions in a bunch crossing is distributed according to Poisson distribution:
\begin{linenomath}
\begin{align}
p_k = \frac{\mu^k}{k!} \frac{\text{e}^{-\mu}}{1-\text{e}^{-\mu}},
\label{eq:poisson}
\end{align}
\end{linenomath}
where $\mu$ is the mean number of collisions and the term $1-\text{e}^{-\mu}$ appears in the denominator, since $k \geq 1$ because of the minimum bias trigger.

The total ZDC energy is distributed according to the $f(E)$ probability density function, which is expressed using the total probability theorem as
\begin{linenomath}
\begin{align}
f(E) &= g(E) \, p_1 + (g * g)(E) \, p_2 + (g * g * g)(E) \, p_3 + \dots,
\end{align}
\end{linenomath}
where $g(E)$ the probability density function of the energy deposit in a single collision and $*$ denotes convolution.

Taking the Fourier transform of both sides:
\begin{linenomath}
\begin{align}
F(\omega) &= \sum_{k=1}^{\infty} p_k \, G^k(\omega) =  \frac{\text{e}^{-\mu}}{1-\text{e}^{-\mu}} \sum_{k=1}^{\infty} \frac{(\mu \, G(\omega))^k}{k!} =  \frac{\text{e}^{-\mu}}{1-\text{e}^{-\mu}} \left(\text{e}^{\mu G(\omega)} - 1\right),
\end{align}
\end{linenomath}
where $F(\omega)$ and $G(\omega)$ are the Fourier transform of $f(E)$ and $g(E)$ respectively. After expressing $G(\omega)$, $g(E)$ can be written as
\begin{linenomath}
\begin{align}
g(E) = \mathfrak{F}^{-1} \left[ \frac{1}{\mu} \log\left[1+(\text{e}^{\mu}-1) F(\omega)\right] \right].
\label{eq:pu2}
\end{align}
\end{linenomath}

\begin{figure}[t]
\centering
\includegraphics[width=0.32\textwidth]{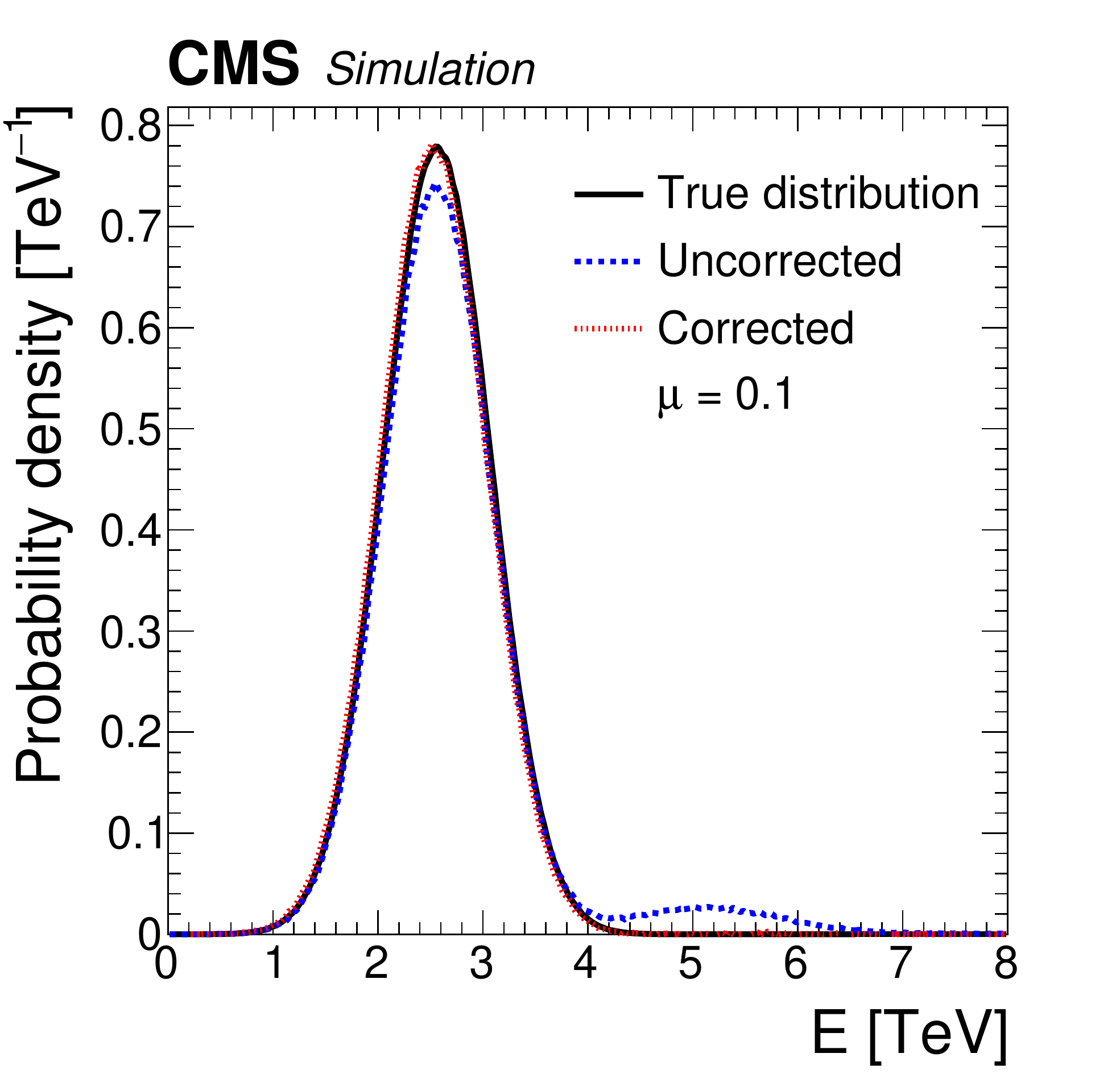}
\includegraphics[width=0.32\textwidth]{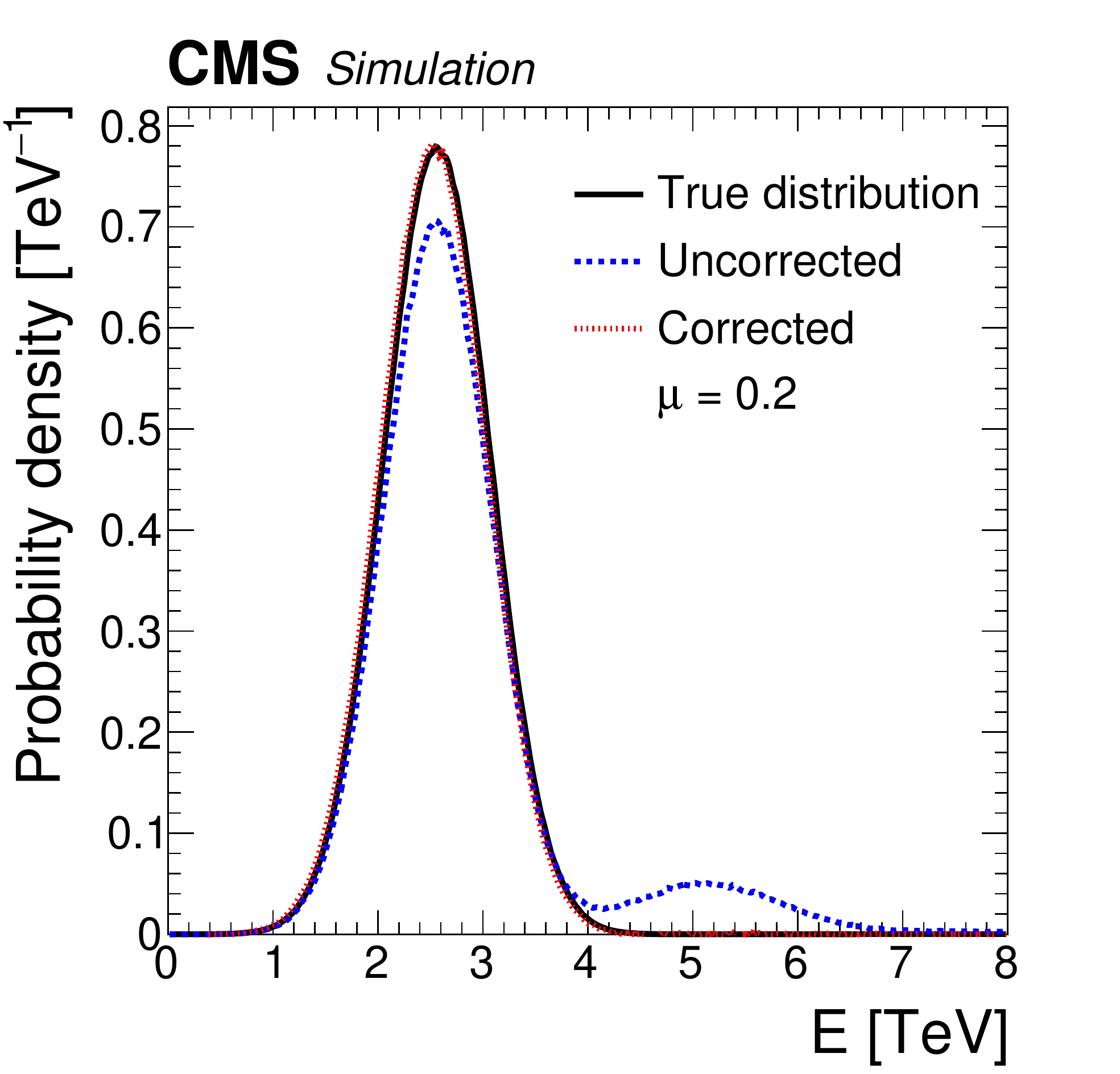}
\includegraphics[width=0.32\textwidth]{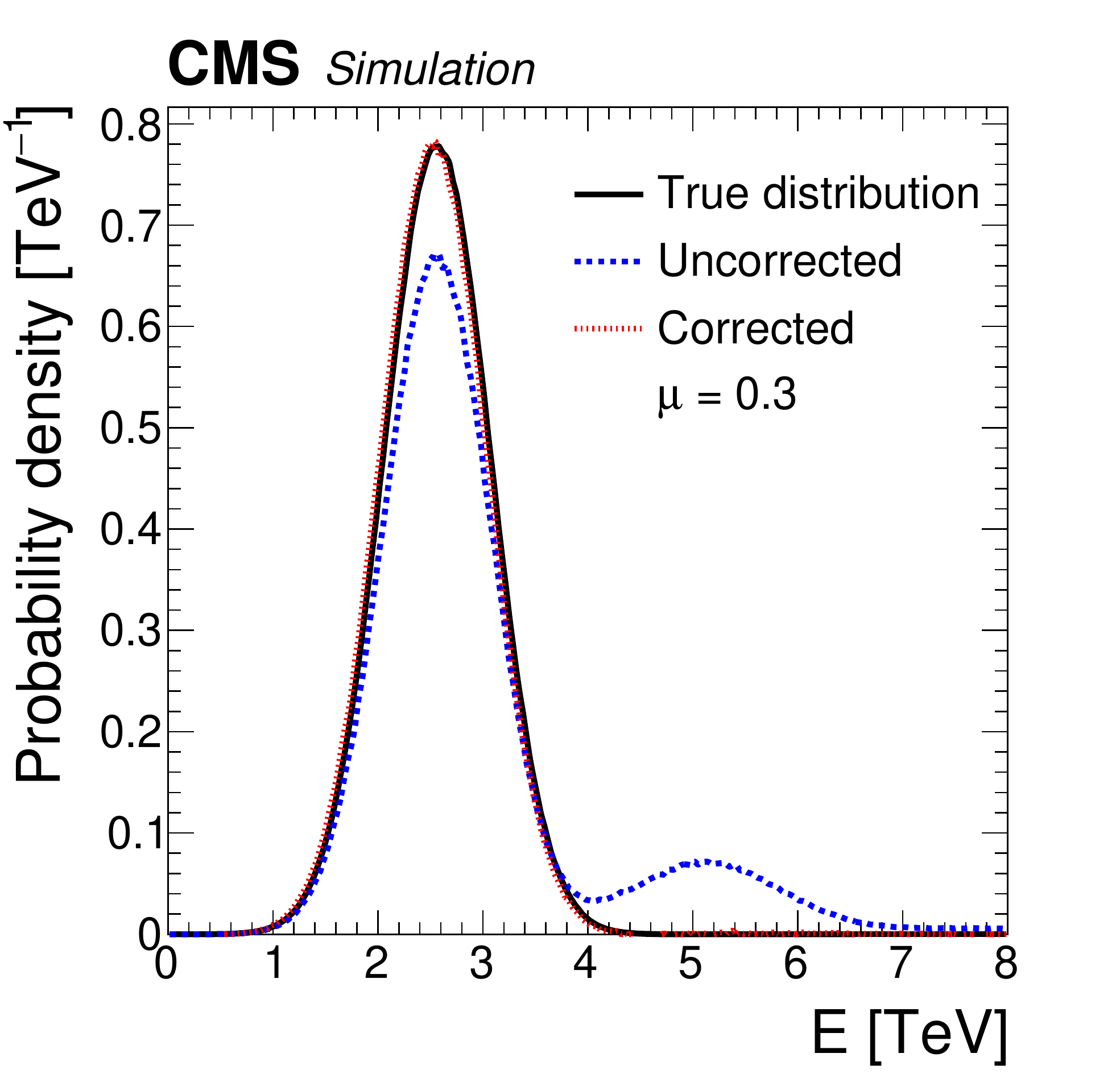}
\caption{Testing pileup correction method on a toy model assuming Gaussian ZDC energy distribution with different pileup values.}
\label{fig:toy}
\end{figure}

The method is tested with a simple model, assuming that the true ZDC energy distribution is Gaussian. First a Poisson distributed random integer $k$ is generated. In the next step $k$ random Gaussian variables are summed. The distribution generated in this way is displayed by the blue curve in Fig.\ \ref{fig:toy}. Finally the Fourier deconvolution is applied to this distribution, and the result (red curve) shows a good match with the true distribution (black curve), supporting the method. This test is performed with various $\mu$ pileup values.

\begin{figure}[t]
\includegraphics[width=0.49\textwidth]{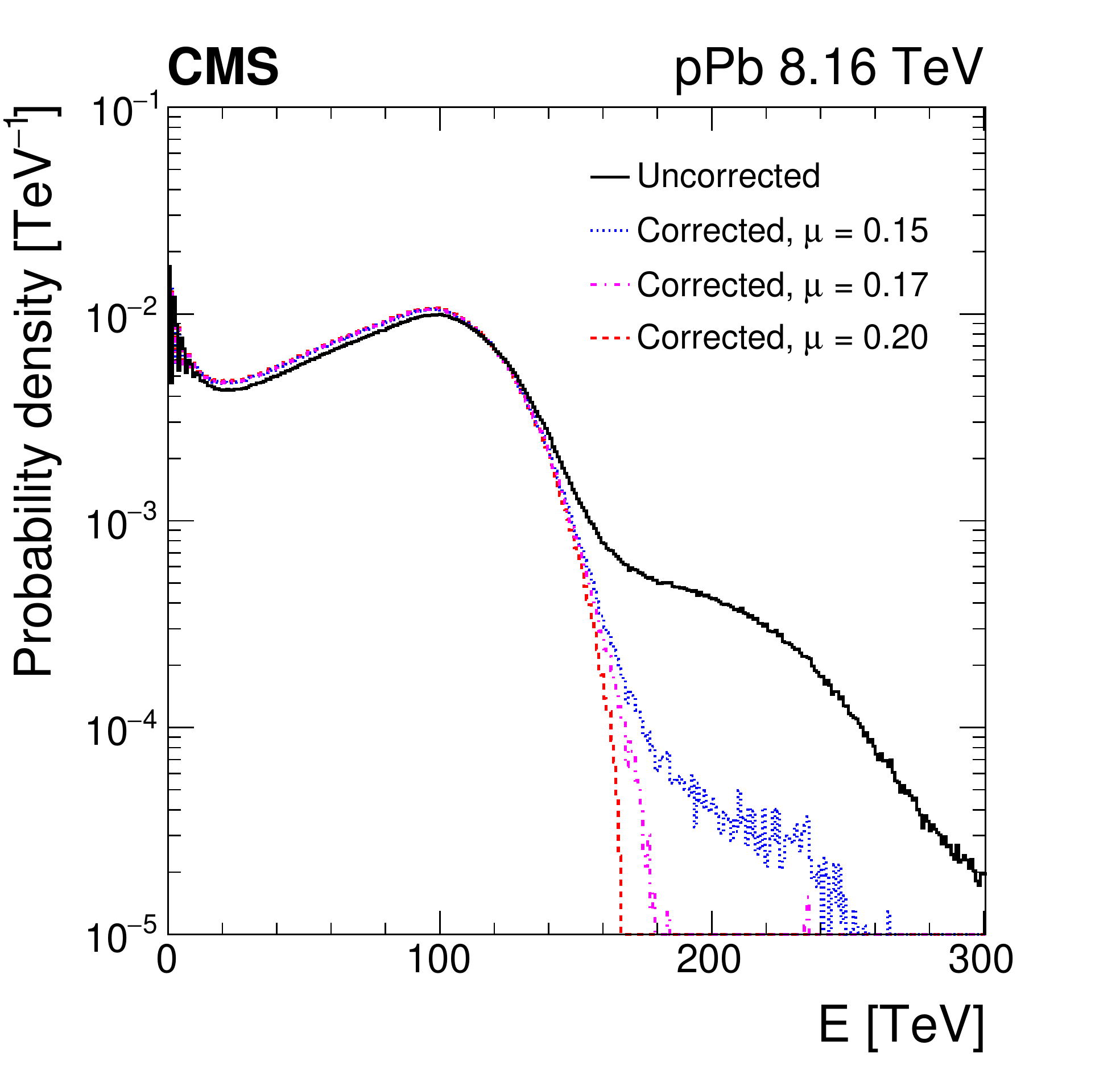}
\includegraphics[width=0.49\textwidth]{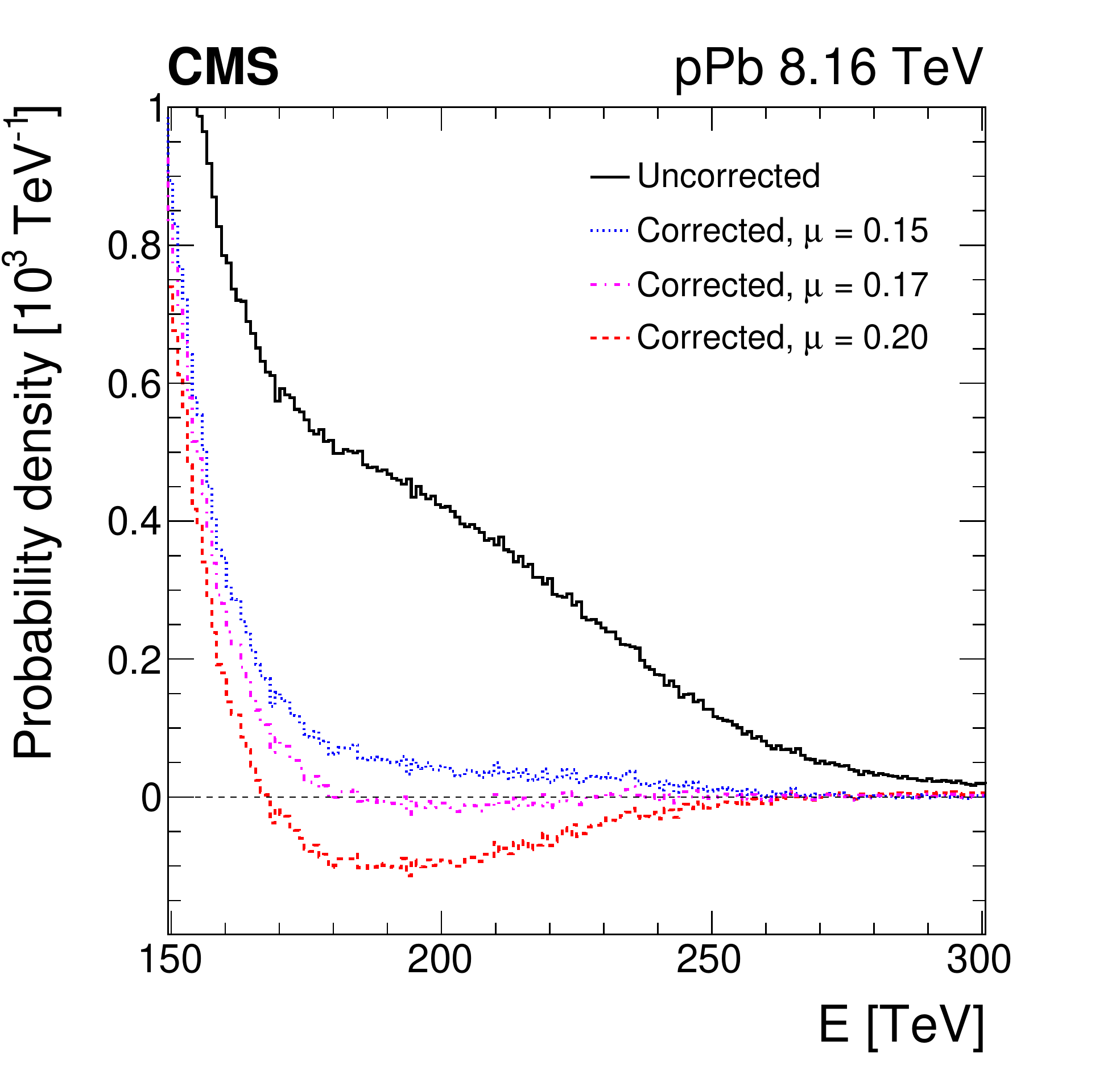}
\caption{Pileup correction applied on data by assuming various $\mu$ values (left). By observing the tail of the pileup corrected distributions, upper limit on $\mu$ can be determined (right).}
\label{fig:pileup}
\end{figure}

The correction is applied to the measured data, the result is shown in Fig.\ \ref{fig:pileup}. As a systematic study, several $\mu$ values are used to perform the correction. The plot in the right panel of Fig.\ \ref{fig:pileup} shows that choosing a too high $\mu$ value in the calculation results in a nonphysical, negative probability density function -- due to the overcompensation of the tail. This provides a possibility to set an upper limit on the value of $\mu$, in our case it is approximately $0.17$. Furthermore, one may obtain a lower limit on $\mu$ as well, from the instantaneous luminosity measured by the central detectors. These luminosity measurements do not include those nuclear excitation processes, when ions are excited and emitting neutrons, but no signal is produced in the central CMS detector.

\section{Conclusion}
The performance studies of CMS ZDC detector have been presented. The response of the detector to neutrons originating from various physics processes was studied using a \textsc{geant}~4 based Monte Carlo simulation, also taking the beam properties into consideration. According to the simulation, a different signal is produced in the electromagnetic and hadronic sections of the ZDC because of the different sampling ratios, thus a weighting factor was introduced to account for this effect. It was found that the theoretical maximum of energy resolution is $17.1\%$ for 2.56~TeV neutrons. Furthermore, the ZDC has greater than $98\%$ geometrical acceptance for neutrons produced in giant dipole resonance, evaporation and cascade processes.

Then a template fitting approach was presented, which is used to extract the signal amplitudes for the individual channels. This method is based on solving a linear system of equations and also includes the treatments of uncertainties and correlations of the pedestal, the digitization and the template shapes. It provides an opportunity to extract signals from events with a pre-pileup signal without introducing a bias.

The channels were gain matched by comparing to the Monte Carlo simulation of the detector and using various data-based techniques. Peaks were observed in the ZDC energy spectrum, corresponding to single, double, and triple neutron events. It was shown, that the spectrum can be described by the sum of two exponential functions, describing the noise peak and photons, and the sum of Gaussian distributions, describing the neutron peaks. It was found, that the parameters of the neutron peaks vary by time, because of the change in instantaneous luminosity. For this effect a simple, event-by-event correction factor was introduced.

Finally, a method using Fourier transformation was presented to correct for the effect of in-time pileup. The feasibility of this correction was demonstrated using a Gaussian toy model. It was shown that by examining the tail of the corrected distribution, an upper limit can be derived on the value of pileup.

\section*{Acknowledgments}
We congratulate our colleagues in the CERN accelerator departments for the excellent performance of the LHC and thank the technical and administrative staffs at CERN and at other CMS institutes for their contributions to the success of the CMS effort. In addition, we gratefully acknowledge the computing centres and personnel of the Worldwide LHC Computing Grid and other centres for delivering so effectively the computing infrastructure essential to our analyses. Finally, we acknowledge the enduring support for the construction and operation of the LHC, the CMS detector, and the supporting computing infrastructure provided by the funding agencies.

The CMS Zero Degree Calorimeter detector is supported by the Office of Science, US Department of Energy. This research is supported by the \'UNKP-19-3 New National Excellence Program of the Ministry for Innovation and Technology, the National Research, Development and Innovation Office of Hungary (K~124845, K~128713, K~128786, and FK~123842), and the Hungarian Academy of Sciences ''Lend\"ulet'' (Momentum) Program (LP~2015-7/2015).

\bibliographystyle{JHEP}
\bibliography{references}

\end{document}